\documentclass[acmsmall]{acmart}

\usepackage{graphicx}
\usepackage{ulem}
\usepackage{booktabs} 
\usepackage{multirow}

\usepackage{multicol}

\usepackage{pdfpages}

\usepackage{color}
\usepackage{amsmath}
\usepackage{latexsym}
\usepackage{geometry}
\usepackage[english]{babel}
\usepackage{blindtext}
\usepackage{graphicx}
\usepackage{subfigure}
\usepackage{indentfirst}
\usepackage{ulem}
\usepackage{float}
\usepackage{indentfirst}
\usepackage{setspace}
\usepackage{makecell}
\graphicspath{{./figure/eps/}}

\newcommand{\added}[1]{\textcolor{black}{#1}}
\newcommand{\modif}[1]{\textcolor[rgb]{0.5,0.5,0.9}{#1}}
\newcommand{\old}[1]{\textcolor[rgb]{0.5,0.5,0.5}{OLD: #1}}

\newcommand{\newcl}[1]{\textcolor{orange}{CL:#1}}

\renewcommand{\modif}[1]{#1}
\renewcommand{\newcl}[1]{#1}
\renewcommand{\old}[1]{}

\newcommand{\hlight}[1]{\textbf{#1}}
\newcommand{\quoted}[1]{{\it{``#1''}}}

\newcommand{\sy}[1]{\textcolor{black}{#1}}

\newcommand{\modality}[0]{{{\sc Communication Modality}}}

\newcommand{\ef}[0]{{{\it Audio only}}}

\newcommand{\ws}[0]{{{\it Audio+Screen}}}

\newcommand{\rauthor}[0]{{Author}}
\newcommand{\rtypist}[0]{{Typist}}

\AtBeginDocument{%
  \providecommand\BibTeX{{%
    \normalfont B\kern-0.5em{\scshape i\kern-0.25em b}\kern-0.8em\TeX}}}

\setcopyright{acmlicensed}
\acmJournal{PACMHCI}
\acmYear{2022} \acmVolume{6} \acmNumber{CSCW2} \acmArticle{338} \acmMonth{11} \acmPrice{15.00}\acmDOI{10.1145/3555758}

\begin{document}

\title{The Name of the Title is Hope}

\title[Typist Experiment]{Typist Experiment: \newcl{an Investigation of Human-to-Human Dictation via Role-play to Inform Voice-based Text Authoring}}


\author{Can Liu}
\authornote{Corresponding Author}
\email{canliu@cityu.edu.hk}
\affiliation{%
  \institution{School of Creative Media, City University of Hong Kong}
  \country{Hong Kong SAR, China}
}

\author{Siying Hu}
\email{siyinghu-c@my.cityu.edu.hk}
\authornote{Student first author}
\affiliation{%
  \institution{School of Creative Media, City University of Hong Kong}
  \country{Hong Kong SAR, China}
}

\author{Li Feng}
\email{feliciafeng35@gmail.com}
\affiliation{%
  \institution{School of Creative Media, City University of Hong Kong}
  \country{Hong Kong SAR, China}
  }


\author{Mingming Fan}
\affiliation{%
  \institution{Computational Media and Arts Thrust, The Hong Kong University of Science and Technology (Guangzhou)}
  \city{Guangzhou}
  \country{China}
}
\affiliation{%
  \institution{Division of Integrative Systems, Department of Computer Science and Engineering, The Hong Kong University of Science and Technology}
  \country{Hong Kong SAR, China}
}
\email{mingmingfan@ust.hk}

\renewcommand{\shortauthors}{Can Liu et al.}

\begin{abstract}


Voice dictation is increasingly used for text entry, especially in mobile scenarios. However, the speech-based experience gets disrupted when users must go back to a screen and keyboard to review and edit the text. While existing dictation systems focus on improving transcription and error correction, little is known about how to support speech input for the entire text creation process, including composition, reviewing and editing. We conducted an experiment in which ten pairs of participants took on the roles of authors and typists to work on a text authoring task. By analysing the natural language patterns of both authors and typists, we identified new challenges and opportunities for the design of future dictation interfaces, including the ambiguity of human dictation, the differences between audio-only and with screen, and various passive and active assistance that can potentially be provided by future systems. 

\end{abstract}

\ccsdesc[500]{Human-centered computing~Empirical studies in HCI}

\keywords{dictation, speech, text input, authoring, role-play, intelligent interface} 

\received{April 2021}
\received[revised]{November 2021}
\received[accepted]{March 2022}

\maketitle
\section{Introduction}
The great potential of voice interfaces has been widely recognized over decades, with research showing that text input via speech is much faster than typing \cite{10.1145/3313831.3376861,Karat:1999:PEC:302979.303160,10.1145/371127.371166}. 
Recent breakthroughs in Speech Recognition~\cite{Hinton2012,sak2014long} 
and Natural Language Processing (NLP)~\cite{chiu2018} 
have dramatically improved the ability of machine intelligence to understand speech~\cite{Hinton2012,chiu2018}. Nowadays, Speech recognition is available on most mobile devices, yet speech is far from being as widely accepted as typing for text input and even the state of the art of voice-based text authoring still faces limited usage. Existing dictation systems focus on transcription and error correction~\cite{suhm1997exploiting,karat1999patterns,ogata2005speech}. There is a lack of support for the entire loop of interaction of text input, including text composition, review and editing~\cite{10.1145/3313831.3376861,10.1145/3173574.3173977,10.1145/3313831.3376173}. With existing dictation systems,  users start by generating text by speaking, but then typically have to fall back to the keyboard for reviewing and editing it~\cite{10.1145/3173574.3173977}. This switch of modality interrupts the voice-based experience, leading to breakdowns especially in mobile scenarios when people are walking, cooking or driving~\cite{10.1145/3313831.3376173}.

To alleviate this problem, future dictation interfaces need to not only improve the accuracy of transcription, but also support a more well-rounded experience of text authoring via voice. While it is almost impossible to write an essay purely using voice today with existing dictation systems, people were writing books in the old time with professional typists. How can we design dictation systems that can support all the fundamental tasks in authoring a piece of text, including composing, reviewing and editing? 
To inform the design of such interfaces, it is important to have a holistic and in-depth understanding of users' natural speech patterns for doing that.

One commonly used method to study natural user behaviors for designing future systems is Wizard-of-Oz~\cite{8155004}. Wizard-of-Oz studies use human to simulate a system in order to evaluate the user experience before developing a functional system that is intelligent enough to understand free speech and distinguish different modes of operations. In this case the simulated machines are often constrained by only responding to a predefined set of commands and rules given to the participants, thus the results of the studies are limited by technical constraints and design choices made by experimenters. On the other hand, existing studies for understanding natural speech focus on improving speech recognition by addressing problems caused by disfluencies~\cite{liu2006enriching,kourkounakis2020detecting} and how users address speech recognition errors by changing the way they speak~\cite{10.1145/371127.371166,10.1145/1216295.1216339}. Therefore it remains unclear how users would naturally use dictation to author an entire piece of text. 

To fill this gap, we adopt role-play as a new method for studying natural dictation. Previously role-play has been used as a design method for experiencing low-fidelity prototypes ~\cite{simsarian2003take, svanaes2004putting} and a teaching method~\cite{Moroz2009role, toms1985effective} for engaging students. This work invents a new use for it: by learning from how humans naturally perform a task with the assistance of another human, we generate insights and inspirations for the design of natural user interfaces potentially powered by machine intelligence.
We conducted an experiment in which participants dictated text to another human, akin to speaking to a typist in the old time. By analyzing the natural speech patterns of both the authors and the typists as well as how they interacted with each other, we aimed to derive insights and opportunities for designing future dictation interfaces for text authoring. 
In the experiment, 10 pairs of participants took the roles of \sy{\emph{Authors} and \emph{Typists}} to perform tasks of composing and editing text until it is polished and ready for publication. 
Afterwards, Authors and Typists switched their roles for the second half of the experiment. 
In addition, as dictation interfaces are used in both eyes-free or with-screen situations, we tested both conditions in our experiment and investigated the differences. 

The main contributions of this work are the empirical findings and design implications, generated from the quantitative and qualitative analysis of natural speech patterns in human-to-human dictation. 
\newcl{Specifically, we extracted a comprehensive list of behavior patterns from the Authors and Typists. We found the Authors composed and edited text by instructing the Typists with the following behavior patterns: \textit{creating new content, re-speaking, explicitly locating and editing text, reviewing text, delegating task} and \textit{thinking aloud}. Meanwhile, the Typists provided the Authors with the following verbal assistance alongside typing: \textit{passively responding to requests}, \textit{actively correcting and preventing errors, proposing reviews or ideas}, and \textit{taking over by making unsolicited edits}. Among these, we discovered how \textit{re-speaking} was used for \sy{five} different purposes implicitly by the Authors and \sy{two} other purposes by the Typists. Moreover, we identified their deictic references and communicating strategies used for locating text and resolving misunderstandings, as well as how the Authors and Typists synchronized and adapted to each other quickly throughout the experiment. In addition, we found that while seeing the text made it easier for \sy{Authors} to review and edit efficiently, not seeing the text was sometimes preferred as it let their thoughts flow freely without distraction.}

\newcl{Beyond the particulars, we are the first to use a role-play method to understand how users could naturally interact with another human, who acts as a service provider comparable to an intelligent system, to inform the design of future interfaces. Despite the likely differences between human-human and human-machine interaction, our findings provides inspirations for system designers and engineers by unpacking new understandings of natural user behaviors, which can be potentially supported by future systems.}

\section{Related Work}
Our work was inspired and informed by related works in three areas: Voice-based Dictation for User Interfaces, Understanding Natural Speech Patterns, and \newcl{the use of role-play method}. 

\subsection{Voice-based Dictation User Interfaces}

Automatic speech recognition (ASR) based dictation has been studied for different tasks (e.g., composing and transcribing) and for different target user groups (e.g., people with learning disabilities~\cite{de1999composing}, blind users~\cite{azenkot2013exploring}, or doctors and nurses~\cite{pezzullo2008voice,viitanen2009redesigning,carter2009voice}). \newcl{Previous work found that blind users used speech input more often than sighted users on mobile devices~\cite{azenkot2013exploring}. Their study showed that blind users who perform text input with speech could do it 5 times faster than with a keyboard, but editing recognition errors was frustrating and it took 80\% of the task completion time. 
Another research on the use of speech recognition software, which studied user groups with and without physical disabilities, found that on average, users could spend around 66\% of their time on correcting dictation errors~\cite{sears2001productivity}.}
These findings show that dictation and correction are two main activities when using ASR-based dictation software, and that correcting dictation errors remains a key challenge. 
This inspired researchers to investigate users' expectations and strategies for correcting dictation errors. Basapur et al. studied users' expectations regarding dictation on mobile devices, finding that users would prefer to correct the errors by using voice commands instead of typing~\cite{suhm2001multimodal}. 
Sear et al. studied voice commands for navigating to the errors in the text and found that the direction-oriented navigation (e.g., move up two lines) was less effective than the target-oriented navigation (e.g., select target)~\cite{sears2003hands}.

Researchers also investigated various interaction techniques to help users correct their errors. In a desktop environment, Suhm et al. showed that multi-modal error correction that combines techniques of respeaking, handwriting, pen-based gestures, and keyboard input, is more efficient than uni-modal error correction~\cite{suhm2001multimodal}. For mobile devices, Kumar et al. designed Voice typing that uses a marking menu with touch gestures for faster error corrections~\cite{kumar2012voice}. Ghosh et al. designed EDITalk to support quick identification of sentence boundaries and speech commands~\cite{10.1145/3313831.3376173}. Recently, Ghosh et al. also designed a technique called VoiceRev to support 2 common types of eye-free editing: commanding and re-dictation~\cite{ghosh2020commanding}. 

Although these interactive techniques enable users to better locate and correct diction problems, they were designed around the current ASR technique, which has little to no ability in understanding users' intents (e.g., composition, correction, or task-unrelated utterances). As ASR continues to improve, we envision that in the near future, it would be able to understand the nuances in users' dictation (e.g., intention, satisfaction) and even have human-like conversations to collaboratively resolve errors in dictation. Towards this future, we seek a more holistic understanding of natural human dictation to inform future dictation interfaces. We achieve this by observing \textit{how the \sy{Author} who dictates, and the \sy{Typist} who writes, can  work collaboratively to compose and edit a given text}.

\subsection{Understanding Natural Speech Patterns}

Many previous studies have observed and analyzed how people speak, either to a speech recognition system or to other humans. Disfluencies, such as self-repair - people correcting their speech immediately after an erroneous phrase is spoken - have been  studied extensively in these literature. By analyzing the characteristics of speech disfluencies, researchers have been trying to identify and correct them automatically. For example, Nakatani and Hirschberg ~\cite{10.3115/981574.981581} created a predictive model - the Repair Interval Model (RIM) that uses lexical, prosodic and acoustic cues - 
to detect and correct self-repair in spontaneous speech. This model has been commonly used in transcription systems ~\cite{kawahara2007intelligent}. Human-to-human speech has also been studied to facilitate speech recognition. For instance, Yang Liu et al. ~\cite{liu2006enriching} studied the repetitions in human conversations, such as ``On Monday I- On Monday I am going to...'', and the use of fillers like ``so'', ``anyway'', ``I mean''. They proposed a computational method to extract such disfluencies in order to correct the speech recognition output to improve its readability. 
Research also found hyper articulation, where user tries to recover from speech recognition error by elongating utterances, pausing, or by altering the pitch. Furthermore, Large et al.~\cite{large2017steering} suggested that people speak to intelligent agents as if they were human. When tasked to speak to a simulated driving assistant, users were observed to be using polite words, deictic references, as well as giving vague instructions. \newcl{However, these previous works focused on analyzing speech patterns to either improve speech recognition accuracy or to inform the design of a conversational agent. Their goals were not for generating texts.}

\newcl{In a previously mentioned work, VoiceRev, the authors conducted a Wizard-of-Oz study to observe users' eyes-free speech patterns while composing and editing text via a human-simulated voice interface~\cite{ghosh2020commanding}. Although some of the behaviors observed in their study also appeared in our work, their study focused on identifying editing strategies that can be immediately implemented, including the use of commands and respeaking. 
While Wizard-of-Oz studies can simulate some intelligent system behaviors, they are limited by predefined interface and functions, and their findings are subject to participants' assumptions and biases about what "an intelligent system" is capable of. Our role-play study complements that by unpacking a much richer and interactive process between humans while covering all phases of text authoring including composition, review and editing.}

\subsection{\newcl{The use of role-play method}}
\newcl{Role-play is a simulation technique to deliberately construct an experience under controlled conditions as designed by experimenters or therapists. It was used in social psychology experiments for studying group dynamics, attitude changes, and in clinical uses for therapies~\cite{Moroz2009role}. 
In HCI, role-play has been mainly used as a design method, particularly for early stage prototyping. 
Blinder~\cite{binder1999setting} 
used role-play with low-fidelity prototypes to get users involved in the design of a PDA-based system. Vogiazou et al.~\cite{vogiazou2007use} used this method to understand which features a new IT system needs by acting out their standard work processes.
Simsarian described how role-play was used at the design company IDEO, where they let clients and end-users assume various roles in bodystorming activities~\cite{simsarian2003take}. 
Howard et al.~\cite{howard2002using} 
introduced professional actors in scenario-based participatory design to enhance immersion. 
Brandt and Grunnet~\cite{brandt2000evoking} 
explored using elements in drama, such as settings, scenarios and props, in a collaborative user-centered design process.
Svanaes et al.~\cite{svanaes2004putting} developed a theoretical framework and a format of process for running workshops with role playing and low-fidelity prototyping, to allow simultaneous exploration of future technology use and design. 
Seland had system designers assessing the role-play method, who found it beneficial for active participation of end users, faster ideation in the early stages of product design, 
and for enhancing developers' understanding of the context-of-use~\cite{10.1145/1182475.1182499}. 
More recently, Buruk and Özcan added wearable devices in role-play games to facilitate movement-based play in game research~\cite{10.1145/2994310.2994315}.} 
\newcl{Furturemore, role-play is also being used as a teaching method in system development courses. For instance, Moroz-Lapin and Maxim~\cite{maxim2017use,Moroz2009role} asked students to act as potential users of a system to enhance their understanding of the requirements as well as the use of the system. 
In addition, Stokoe used role-play for communication skills training  
across a number of workplace settings~\cite{stokoe2011simulated}.} 

\newcl{
All the above mentioned uses of role-play methods focused on engaging users or stakeholders in early stages of participatory design processes in order to create immersion or empathy. In this work, we create a novel use of a different role-play method for designing future systems, with a particular focus on observing natural human behaviors in interacting with an “intelligent party”, which in our case is another human who is not hiding behind the “wizard's” curtain. While we acknowledge the differences between human-to-human interactions and human-computer interactions, this approach aims to unpack unknown natural behaviors of users and to provide new inspirations for future intelligent systems.} 

\section{The Typist Experiment}

\newcl{The major advantages of dictation include the high speed of text input and the possibility to perform dictation eyes-free and hands-free. 
This makes its use scenarios on mobile devices very compelling, where staring at a screen is inconvenient and typing on a keyboard is relatively slow. In such scenarios, text of different lengths need to be created, from short to long messages, emails, memos, diaries, blogs, etc. In this work, we seek to understand how users might use natural dictation to compose and edit whole paragraphs of text. 
We designed a role-play experiment -- the Typist Experiment, to achieve this by observing natural human-to-human interaction, in which one participant (i.e., the Typist) provides a ``smart dictation service'' by typing what the other participant (i.e., the Author) dictates to him/her.}

\newcl{While speech is the main modality of input for dictation, users rely on visual or auditory feedback to understand how the task is being executed. Therefore we believe one important factor that would affect users' behavior and experience is the modality of feedback.  
In fact, using dictation software eyes-free is compelling for many use cases especially in the mobile scenarios, such as while walking, driving or doing other tasks. In order to understand the differences between eyes-free and seeing the text, we make modality as a main independent factor in this experiment.} With it, we sought to answer the following research questions (RQs):
    \begin{itemize}
        \item \textbf{RQ1}: How do Authors dictate to Typists?
        \item \textbf{RQ2}: How do Typists assist Authors? 
         \item \textbf{RQ3}: How do Authors and Typists coordinate and collaborate? 
        \item \newcl{\textbf{RQ4}: How does the communication modality (\ef{} and \ws{}) affect Authors, Typists and their cooperation?}
    \end{itemize}


\subsection{Experiment Design}

The experiment followed a within-subject design featuring one main factor with two independent variables: \modality{} [\ef{}, \ws{}]. 
Pairs of participants were recruited and asked to co-create written text on given topics, each taking the role of an \rauthor{} and a \rtypist{} respectively.
In \ws{} condition, the \rtypist{} shared his/her screen showing the editor interface where text was being typed in and the real-time word count was shown. Whereas in \ef{} condition the screen was not shared, and the \rauthor{} could not see the text. Cameras were all off, to keep speech as the only communication modality besides the editor interface in \ws{} condition. Each participant took the role of \rauthor{} or \rtypist{} to begin with, and then switched roles with their partner. Thus half of the participants took the role of \rauthor{} first and the other half were \rtypist{} first. We expected that the experience of being \rauthor{} and \rtypist{} may influence their behaviors in their second role. 
This was an intentional choice to collect more data and potentially observe richer behavior patterns in their collaboration strategies.

\begin{figure}
  \centering
  \includegraphics[width=0.6\linewidth]{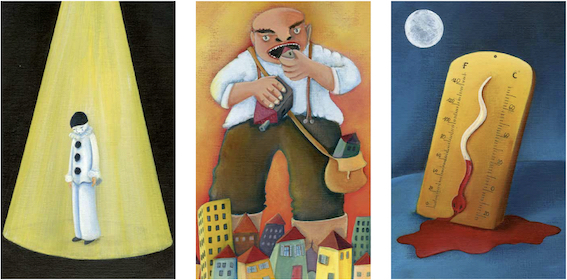}
  \caption{Example images provided for participants for text composition, from the Dixit game.}
  \Description{Example images provided for participants for text composition, from Dixit game.}
  \label{fig:ExprimentPic}
\end{figure}

As an \rauthor{}, the participant was asked to compose a piece of text by describing a given image. The image (Fig.~\ref{fig:ExprimentPic}) is randomly chosen from an image set of the Dixit game\footnote{https://www.libellud.com/}, which is an ambiguity board game where every image card is an abstract illustration designed for having multiple interpretations. 
In our experiment, Dixit cards were used as the source of inspiration or description basis in authoring tasks. \newcl{This choice was to ensure certain control on the task difficulty without limiting the creative space or introducing biases by particular topics. As the experiment focused on observing the interaction patterns between Authors and Typists, the actual content they composed was left to their choice to ensure they felt comfortable with the task.}
The \rauthor{} was asked to create a text of 70-80 words by speaking to the \sy{\rtypist{}} and instructing him/her to modify the content as the \rauthor{} wanted. \newcl{We chose a paragraph-length text generation task to cover the writing situations in most writing tasks on mobile, such as for messages, emails and memos, without making the experiment too long.} A task was finished when the text reached the requested range of length, free of mistakes and the \rauthor{} was satisfied with it so as to be willing to publish it on social media or send it to friends. 

As a \rtypist{}, the participant needed to type in what the \rauthor{} composed and edit it according to the requests and instructions from the \rauthor{}.
The \rtypist{} cannot see the image given to the \rauthor{}. Both participants were told to communicate freely while trying to complete the task as fast as possible.

\subsection{Apparatus}
The experiment was conducted online, with the \rauthor{}, the \rtypist{} and the experimenter in the same virtual meeting session. Zoom\footnote{https://zoom.us/} or Skype\footnote{https://www.skype.com/en/} or TencentMeeting\footnote{https://www.tencent.com/en-us/responsibility/combat-covid-19-tencent-meeting.html} was used as the meeting platform, considering the participants' preferences and their regional network situations. Standard Google Doc\footnote{https://support.google.com/a/users/answer/9282664?hl=en} or Tencent Doc\footnote{https://docs.qq.com/} was used as the editor, and the real-time word count view was configured to be visible in all conditions. Participants were asked to disable their camera for reasons introduced above. 
\added{Video and audio were recorded using the experimental platform built-in software function (e.g., Zoom and Skype recording) combined with screen recording software.} 

\subsection{Procedure}
Before starting the experiment, the experimenter introduced the study, collected the informed consent and sent designated images to the \rauthor{}. 
\newcl{The roles of the Author and Typist were introduced at the beginning of the experiment, as the Author being the writer and the Typist being the service provider.}
\old{Twenty-four} 
\modif{24} images were provided to each Author (\sy{three} per trial for them to choose \sy{one}) during the entire experiment. Each set of images for a group were randomly selected from a fixed set of 84 Dixit cards. We chose to let participants type in words instead of using speech to text on Google Doc. This was to avoid introducing the voice typing interface as a third player in the interaction, which would distract the interaction between the two parties with a cooperative error correction task caused by the limitation of current STT products. The experimenter checked the function settings of the online typewriting platform used by the \sy{\rtypist{}} before the test started. The document platform turned off all notification-related functions, such as grammatical error display and spelling suggestion function, to provide participants with pure documentation tools. The purpose was to reduce the influence of machine assistance and visual reminders' interference, which helped to observe and record the participants' natural voice communication behaviour in the experiment.
The participants began with a training session before starting each condition to familiarize their role and the task condition. For each trial, the experimenter provided three different images for the \rauthor{} to choose one for composing text. They were told to freely describe the image or talk about experiences or opinions inspired by the image. Images were randomly chosen from the set, and we ensured that different images were provided for each trial of one experiment. 

Each participant was assigned a role as the \rauthor{} or the \rtypist{} to begin with. Each measured trial required the completion of composing and editing one piece of text. Two repetition trials per \modality{} condition were performed in the first round. Then the participants switched their roles and performed another \sy{two} trials per \modality{} condition. \newcl{The order of \modality{} was counterbalanced across participants.} The participants began with a training session before starting each condition to become familiar with their role and the task condition. For each trial, the experimenter provided three different images for the \rauthor{} to choose one for composing text. They were told to freely describe the image or talk about experiences or opinions inspired by the image. \newcl{Each participant completed \sy{four} trials as Author and \sy{four} other trials as Typist in total. The participants were strangers to each other, and were randomly paired. There was no restriction on communication, but participants were asked to complete the task as fast as possible while ensuring a sufficient quality.}

The experiment in total collected 2 \modality{ $\times$ 2 roles $\times$ 2 repetition $\times$ 10 groups = 80 measured trials.} After finishing all the trials, we conducted a semi-structured interview with each group. The experiment took around two hours for each group, with a break in the middle. 

\begin{table}[h]
    \centering
    \resizebox{0.5\textwidth}{!}{%
    \begin{tabular}{ccccc}
    \hline
    \textbf{Participant} &
      \textbf{Group} &
      \textbf{Gender} &
      \textbf{\begin{tabular}[c]{@{}c@{}}Education/\\ Profession \end{tabular}} &
      \textbf{\begin{tabular}[c]{@{}c@{}}Native    \\ Speaker\end{tabular}}  \\ \hline
    P1  & \multirow{2}{*}{G1}  & F & Postgraduate          & No   \\
    P2  &                      & M & Postgraduate           & No   \\ \hline
    P3  & \multirow{2}{*}{G2}  & M & Postgraduate           & No   \\
    P4  &                      & M & Postgraduate           & No   \\ \hline
    P5  & \multirow{2}{*}{G3}  & M & College Lecturer  & Yes  \\ 
    P6  &                      & F & College Lecturer      & Yes  \\ \hline
    P7  & \multirow{2}{*}{G4}  & F & Undergraduate  & No   \\
    P8  &                      & F & Undergraduate  & No   \\ \hline
    P9  & \multirow{2}{*}{G5}  & F & Postgraduate          & No   \\
    P10 &                      & M & Postgraduate          & No   \\ \hline
    P11 & \multirow{2}{*}{G6}  & F & Postgraduate           & No   \\
    P12 &                      & M & Postgraduate               & No   \\ \hline
    P13 & \multirow{2}{*}{G7}  & M & Undergraduate  & No   \\
    P14 &                      & F & Undergraduate  & No   \\ \hline
    P15 & \multirow{2}{*}{G8}  & M & Undergraduate  & No   \\
    P16 &                      & M & Undergraduate  & No   \\ \hline
    P17 & \multirow{2}{*}{G9}  & F & Postgraduate           & Yes  \\
    P18 &                      & M & Postgraduate           & Yes  \\ \hline
    P19 & \multirow{2}{*}{G10} & M & Postgraduate          & Yes  \\
    P20 &                      & M & Postgraduate           & Yes  \\ \hline
    \end{tabular}}
    \caption{Summary of participants' demographic information. 
    } 
\vspace{-0.8cm}
\end{table}

\subsection{Participants}
We recruited 20 participants (8 females and 12 males) from local universities to form 10 groups \added{(Table 1)}. None of them had professional typewriting or transcription training. To reduce biases of personal relationships, we paired the participants so that each pair did not know each other. The tasks were performed in English. 
\old{The participants were non-native speakers but were fluent in English as registered full-time students in English-taught universities.} 
The participants included three groups of native English speakers and seven groups non-native English speakers. All the non-native English speakers studied in English-taught university programs and a score of at least 6.5 in IELTS tests.
They had diverse backgrounds and education levels, holding bachelor, master and doctoral degrees in communication, arts, finance, law, chemistry and biomedical engineering.
For typewriting the content of the composition task, eight groups used Google Doc, and two groups used Tencent Doc.
\added{Although some of our participants had typewriting capacity issues, we focused on the nature of the speech patterns between the Typist and the Author in dictation-based text authoring practice. We discussed this point about participant's ability in the limitation part.}

\subsection{Data Collection and Analysis}

The following data was collected: 1) screen and audio recordings of the online experiment sessions; 2) audio recordings of the participant interviews; 3) observational notes from the experimenter. 
For the experiment sessions, we transcribed all the conversations between Authors and Typists into text and performed a Thematic Analysis on the utterances. The analysis was done with the aid of the audio and screen recordings as references. Two researchers independently coded four randomly selected trials of data from two groups, which covered 5\% of the entire data set. They discussed their codes to gain a consensus. After rounds of discussion, their codes reached a substantial level of inter-rater reliability (Cohen’s kappa: k = 0.78). Fig~\ref{fig:author_codes} shows a few examples of how the codes were applied.
After that, one of them continued to code the rest of the data.
All interview sessions were transcribed into text manually and coded by two researchers after reaching consensus about the themes. 
During the coding process, the observational notes were also analyzed. 

\section{Findings}
\newcl{This large section elaborates in detail the behavior patterns we identified from the utterances of both the Authors and Typists, as well as their communication and cooperation strategies. While some examples are given in the main text in prose, we supplement in the Appendix~\ref{tab:author_codeutterances} our coding scheme of all transcribed utterances, together with their descriptions and examples. \sy{Each utterance can be coded with one or more codes in the coding scheme.} The rest of this section label our participant IDs in this format: [Group Number]-[Participants ID]-A/T(Author/Typist), i.e., ``G10P20-A''. We add AO (Audio Only) and AS (Audio+Screen) when comparing modalities.}

\subsection{How The Authors dictated to The Typists? (RQ1)}
\begin{figure}[]
  \centering
  \includegraphics[width=1.0\linewidth]{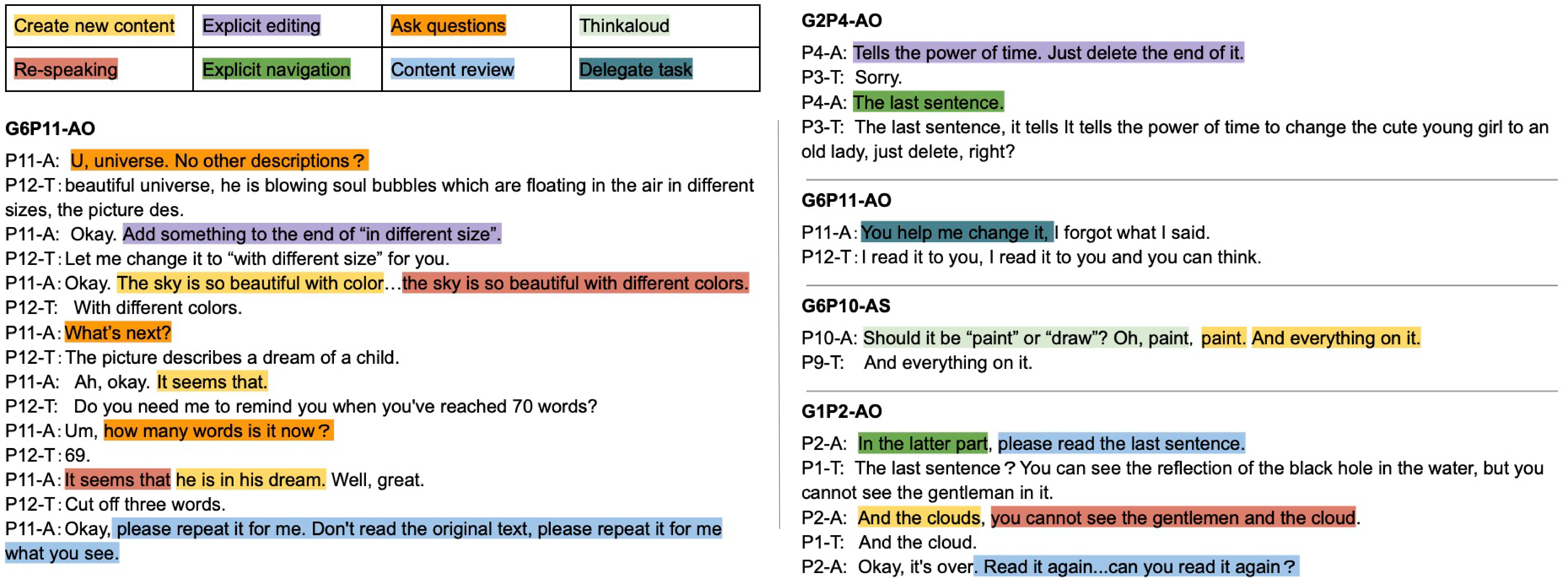}
  \caption{Examples of how Authors' utterances are coded with the 8 categories.}
  \Description{Code transcription based on the utterance category }
  \label{fig:author_codes}
\end{figure}

The analysis of Authors' utterances focus on extracting behavior patterns of the Authors, including how they composed and edited the text by giving implicit and explicit instructions, how they switched between the different types of instructions as well as what other requests they made. The findings presented in this section are based on our categorization of all the utterances from the authors. Fig.~\ref{fig:author_codes} shows examples of how these categories are coded in Authors' utterances.  We calculated the occurrences of each category and normalized the numbers into percentages by dividing them with the sum of coded Author utterances for each group, which gives us an estimation of frequency of each behavior pattern. 
Seven categories emerged in total: Create new content (44.22\%), Re-speaking (37.45\%), Explicit editing (6.06\%), Explicit navigation (2.15\%), Content review (1.76\%), Ask questions (4.52\%), Delegate task (0.64\%) and Think aloud (3.20\%).
The numbers here are average percentages among 20 Authors, and indicate that the primary requests made by Authors were Creating new content and Re-speaking. 

\subsubsection{Ambiguity between composition and edits}

\paragraph{Creating new content}
We use \emph{New Content} as the code for composition of new text. It is coded in chunks of words within utterances. One utterance often includes both newly composed words and previously composed words, which are repeated with or without modifications.

\paragraph{Re-speaking}
\label{sec:respeaking}
Re-speaking is coded as \sy{Authors} repeating chunks of content after it being composed. Existing literature studied speech repair~\cite{10.3115/1034678.1034742}, which we see as one type of re-speaking behavior, typically observed as a sort of stuttering. In our study, we identified \sy{five} types of re-speaking behaviors from the Authors, depending on their different intent. 
\begin{itemize}
    \item \emph{Overwrite to modify}: re-speaking the text to overwrite the different part of it, as an implicit request to modify the text. \added{Example: \quoted{G5P9-A: The \hlight{haircut}...oh...the \hlight{hair style}.} and Example 7 in Appendix A shows the longer overwrite.}

    \item \emph{Confirm or repeat for Typist}: re-speaking to make sure the \sy{Typist} clearly understood and took down what they composed. Example: \quoted{G5P9-A: En...sorry. When we were little child. When we were little child.}
    \item \emph{Continue composition}: re-speaking a few words as a way to continue composing after it, so that the Typist knows where to continue typing. \added{ Example: \quoted{G8P16-A: \hlight{Everyone}, no, no, a new sentence, \hlight{everyone} was telling her.}}
    
    \item \emph{Locate / refer to text position}: calling keywords as a deictic reference to communicate a text position, for making a request then. \added{Example: \quoted{G6P11-A: Front, back to front, back to top. \hlight{`boy'} is it? a \hlight{`boy'} walking on the grass.}}

    \item \old{\emph{Nature speech repair}} \modif{\emph{Natural speech repair}}: repeating words in subconscious stuttering for repairing one's own speech. Example: \quoted{G1P2-A: Is \hlight{the, the} paper man becomes loosening.}
\end{itemize}

The average percentages of each type of Re-speaking in all the Re-speaking utterances are: Confirm/repeat for \sy{Typist} (42.80\%), Overwrite to modify (22.25\%), Continue composition/ Anchor new content (20.05\%), Natural speech repair (6.96\%), Locate/ Refer to previous content (5.44\%).

\paragraph{Explicit editing}

The previous section introduced Re-speaking as a major way of making implicit editing requests. The other way Authors made editing requests to the Typist was to be explicit, closer to what existing dictation software could support, such as the Google Docs Voice Typing\footnote{https://support.google.com/docs/answer/4492226?hl=en}. We observed \sy{six} types of requests: \emph{Add, Delete, Replace, Format, Organize} and \emph{Punctuate}. Different from the voice commands in fixed syntax supported by existing systems, the \sy{Authors'} requests were made in free speech with various syntax. \added{For instance \quoted{G5P10-A: success, no no no no no no} was used to express a ``Delete'' request. Organizing content can be said in many ways, such as, \quoted{G8P16-A: Put all the previous words in a quotation mark. There is one more sentence after the quotation mark. This is Richard.}}

\subsubsection{Mixed composition and edits}

\begin{figure}
  \centering
  \includegraphics[width=1.0\linewidth]{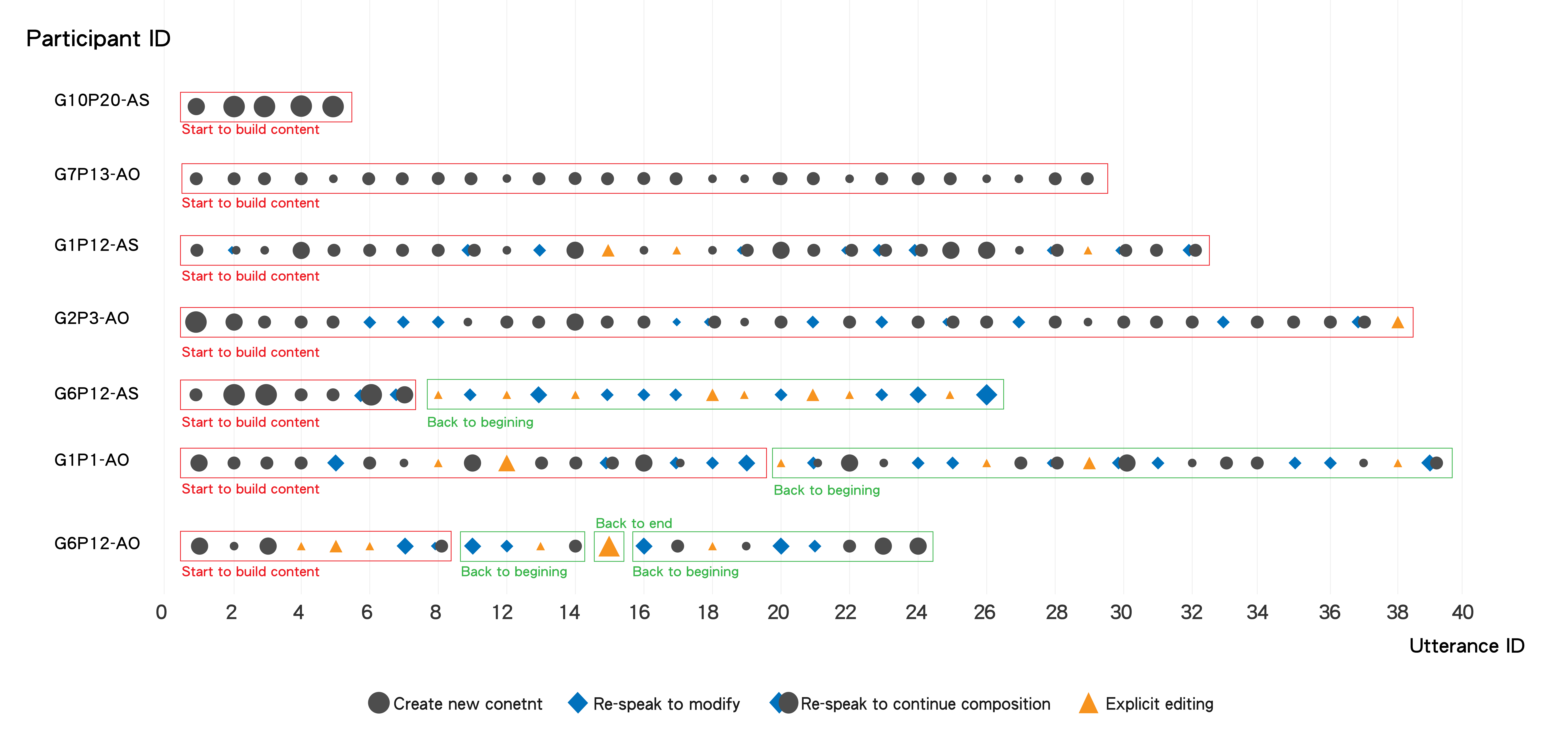}
  \caption{\newcl{Example visualizations of how text got composed and edited over the timeline of a trial. Each symbol represents the content development in \textit{one} utterance: \textit{Create new content, Re-speak to modify, Re-speak to continue composition} and \textit{Explicit editing}. 
  AO (\ef{}) and AS (\ws{}) are communication modalities between Authors and Typists. \sy{Four} sizes of the symbols represent the unit size of text operated in the utterance: \textit{word, phrase, clause} and \textit{multiple clauses}. A red contour contains the first composition pass in the trial and a green contour contains one revision pass.}} 
  \Description{Operations along timeline to visualize the strategies}
  \label{fig:content_strategies}
\end{figure}

Based on the transcriptions, we analyzed how the text evolved over the timeline of each trial. We observed three types of composition strategies. A few examples are visualized in Fig.~\ref{fig:content_strategies} to show this process. There are only \sy{four} operations that affect the text: Create new content, Re-speak to modify, Re-speak to continue composition and Explicit edit. The first strategy was to focus on generating new content first and then go back to the beginning to edit, as seen in a typical example \added{G1P2-AO}. In this case, the tasks were finished after \sy{two} or more passes (see example \added{G1P1-AO, G6P12-AS}). The second strategy was to edit as they compose, the editing instructions were a mixture of explicit and implicit (via Re-speaking) requests. The task was finished with one pass. The third strategy happened rarely with Authors who were able to organize their thoughts and articulate in a smooth sequence in one go without needing to edit (see example \added{G7P13-AO, G10P20-AS)}. 
Every above-mentioned strategy appeared in both \ef{} and \ws{} conditions. These strategies also get mixed up often, leading to a process with less order (example \added{G6P12-AO}). Random jumps of editing locations could cause the Typist to get lost, or needed more synchronization effort.

\begin{figure}
  \centering
  \includegraphics[width=\linewidth]{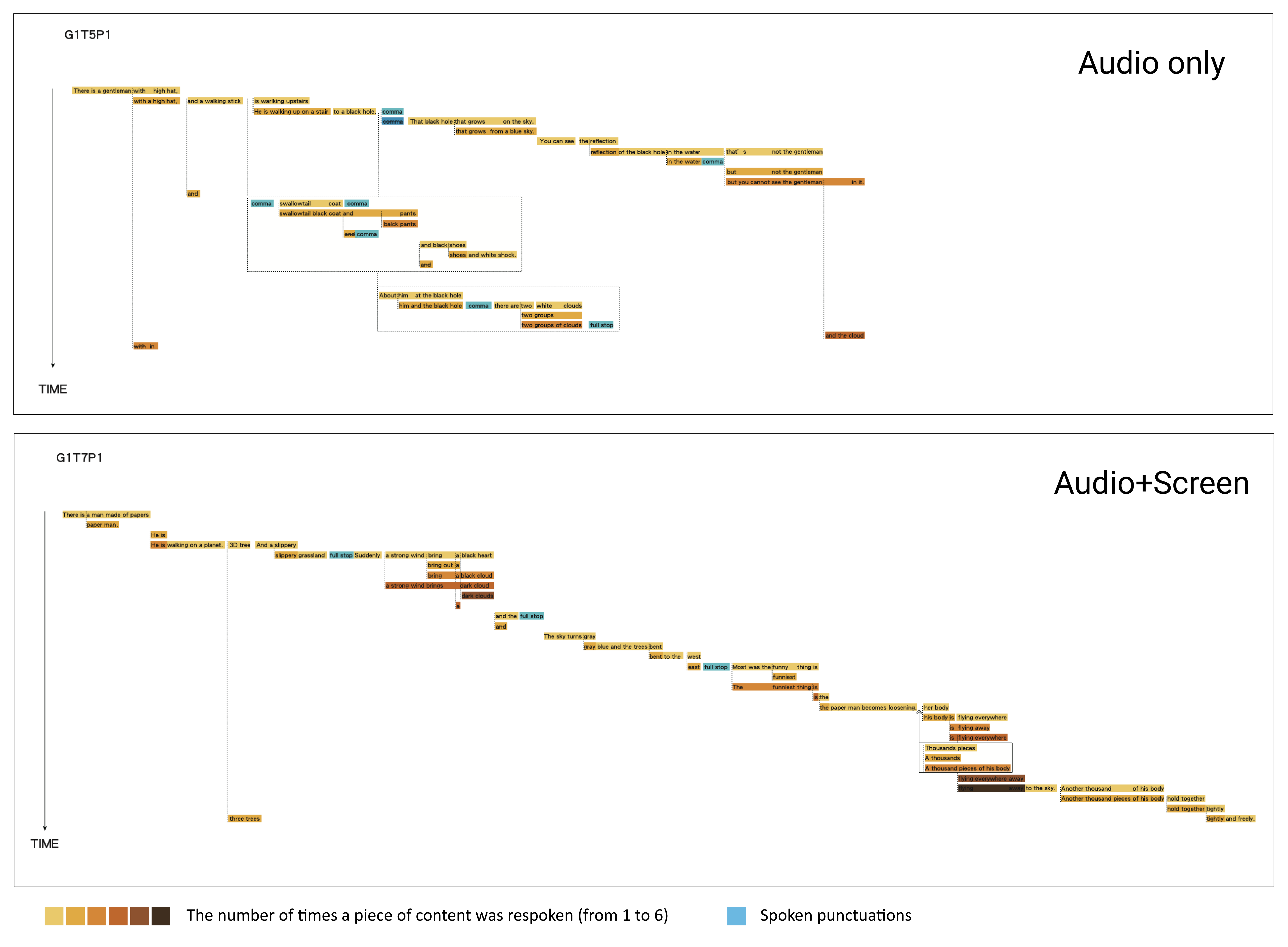}
  \centering
  \caption{\old{Example visualizations of content development process.} \newcl{Example visualizations of the content development process. Overwritten text by re-speaking are aligned vertically across lines. The timeline of text generation runs from left to right and then from the top down. Each coloured text block is generated from one utterance. Darker orange color \sy{indicates} the same content block being \sy{re-spoken} more times. Blue blocks are spoken punctuations. Edits by explicit requests are annotated with crossovers or insertion marks.}}
  \Description{Example visualizations of content development process. }
  \label{fig:content_viz}
\end{figure}

With a more elaborated visualization, Fig.~\ref{fig:content_viz} illustrates in detail how text evolved in time. \newcl{In this figure we ignored other types of utterances from the transcription and focused on \textit{creating new content, re-speaking and explicit editing} as they were the only ones modifying the text. We can see each chunk of text the Author generated or modified by each utterance and over time. Overall we can see many edits done by respeaking, there is no clear mode distinction between composition and editing in nature dictation, nor an easy way to make it. } This example is from G1P1 in \ef{} and \ws{} condition in comparison. This Author did extensive editing in both trials. However the edits were more in order in \ws{} condition, compared to in \ef{} condition there were larger chunks of text being inserted and modified at multiple locations. \newcl{ We provide this visualization for all the trials in the supplementary material of this paper as a data set. From this data set we can see most edits were done in one or multiple passes, with some jumps between editing locations, which happened in both \ef{} and \ws{} conditions. Perhaps this is due to our limited working memory~\cite{baddeley1994developments}, in \ef{} condition, the edits tended to be more extensive after jumping locations. Further research is needed to investigate this.}

\subsubsection{Other requests and behaviors of Authors}

\paragraph{Explicit navigation}
We introduced in the previous section that Authors navigated the text by simply re-speaking a few keywords. In contrast with that implicit navigation, there were also explicit instructions for navigation. For instance, Authors would say: \quoted{first line, at the end (G6P11-A)}, or \quoted{The last sentence, after the last sentence (G2P4-A)} or \quoted{after `black hole' (G1P2-A)}. More about navigation can be found in Section~\ref{sec:between_navigation}

\paragraph{Content review} Content review requests were issued when Authors needed to know what had been written. 
We also noted a small amount of content review requests 
were asking the Typists to summarize the written content instead of reading it. For instance, \added{\quoted{G6P11-A: Tell me what you understood from the writing, don't just read the sentences.}}

\paragraph{Ask questions}
\label{sec:author_ask_ques}
The Authors occasionally asked questions to their Typist. 
The types of questions include: asking how to spell or translate a word; asking for suggestion of words; checking whether the Typist finished typing; checking word counts. \added{For example, an \sy{Author} tracking the typing progress would ask:\quoted{G4P8-A: are you finished?} or \quoted{G5P9-A: So can we move on?}. We observed that some \sy{Authors} ask for suggestions. For instance, \modif{G4P8-A:}\quoted{ Do we need to add something like `Anyway, I’m very so mad' in the end?.}}

\paragraph{Delegate task}
Very occasionally, some Authors asked the Typist to help them with the task when they faced difficulties. They asked the Typist to make the sentence better or compose something new. \added{For instance, \quoted{G6P11-A: Please help me to change it.} or\quoted{G4P7-A: Can you help me think about what else to say?}.} 

\paragraph{Think aloud}
Think aloud has been observed as a natural behavior. For instance Authors spoke their mind when they were unsure about the use of a word, or how to express an idea, or indicating they were taking time to think. This behaviour was also observed in previous Wizard-of-Oz studies~\cite{10.1145/3390889}. 

\subsubsection{Authors' dictation strategies}
\newcl{The participants were asked to describe their dictation strategies in the interview. The following findings describe the conscious efforts they made when being an Author.}  

\paragraph{Consciously reducing uncertainty and repetition} One participant reported she was automatically spelling the words that may introduce confusion or need to double check, \quoted{G1P1-A: When I was the Typist, there were words I wasn't sure about. Then when I became the Author, I would actively spell those words, to avoid repeating it}. She also tried to tell the Typist exactly where she was uncertain, to avoid having to reading back the whole text: \quoted{When I wasn't sure about something, I would pick a sentence for him to read, wouldn't review the whole thing, which is too time-consuming.} 

\paragraph{Like telling stories to children} One participant said he was doing the Author's job pretending he was telling stories to children: making up things quickly and keeping things simple. G3P5-A echo that \quoted{I spoke more slowly than normal. Normally, I speak much faster. I think my strategy was I was imagining that I was telling a story to my niece or nephew. So two little children. So that was my main strategy. Yeah, story, telling two little children. Well, because when you're tying stories to children, you don't have time to edit. And you have to come up with everything very quickly. So you have to keep everything kind of simple. And they always end up asking questions, and you always have to think of something right away. So that's why I think my strategy was pretending like, Yeah, one of like, a little child was asking me to tell a story.}

\paragraph{Formulating thoughts} One participant reported he would formulate his thoughts to make it clear, consistent and avoid major changes. \quoted{G10P20-A: When I was authoring, I was a lot more conscious about what I was going to say. So formulating my thoughts a lot more as opposed to when I'm writing for myself, (...). Because I'm conscious that I need to transfer his information to him. So I want to make sure that it's clear on the one hand. On the other hand, I thought that the exercise was such that you have to put things on paper that are very consistent. And I want to prevent anything like major changes that we need to make to it in order to every time efficient so to speak.}

\paragraph{\newcl{Making orderly speech}} Some Authors explained how they managed to speak in an orderly manner. One \sy{Author} who composed in \sy{one} go without much editing explained that was because he was highly trained in writing and had a very good verbatim memory. 
\newcl{G3P6-A: }\quoted{I think I come from a place of being highly trained in writing (...) I focus more on keeping everything in memory, sort of remembering everything I had said and having a clear picture of everything that I did said so, so that then I didn't need a lot of addition (...). So I basically remembered everything I said.} \newcl{Another participant intentionally slowed down the pace of composition to leave pauses for mentally structuring the whole text. As mentioned by G9P18-A \quoted{And my main strategy for composing was to take pauses between sentences to think. So, it was quite slow. Building up the story in piece by piece slowly was my strategy.}}

\subsubsection{Authors' challenges and their wishlists for Typists}

\paragraph{Feeling rushed and disorganized.} Participants mentioned the difficulty of composing text with voice was the stress and lack of structure. G10P19-A:\quoted{ I think being an \sy{Author} just speaking the story I felt a bit rushed, I guess. Like how it's kind of not panicking, but like trying to think that way the story changed because I was kind of maybe like blurting things out without thinking so thoroughly.} Another participant articulated the challenge in tracking down the non-linear thinking that comes with free speech. G1P2-A: \quoted{You have lots of free space. Maybe you finish one sentence and suddenly thought about another point, but you are still speaking about this point, I wish the Typist could understand. Maybe you have to let him choose one point or note down both and then ask you to confirm your choice. Because you are quite free when you speak. The most difficult thing, is you think about many things and just pour them out randomly.} 

\paragraph{Giving up control.} It was also reported that giving up control to another person was a challenge. G3P5-A:\quoted{ So the main challenge, I think, is having to give up some control. Which I think is fine. It was fun. But you know, it's like the, you know, you get we get so used to typing our own words. }

\paragraph{Communication, rhythm and punctuation.} Further challenges reported of being an Author doing text authoring via a Typist echoed our observations and other parts of the finding, included communication problems caused by homonyms, the need to keep stopping and repeating their speech, how to control the speed of speech, how to express the segmentation of sentences, and how to make Typists understand better the meaning of the sentence.
\added{As G1P2-A echoed \quoted{The main problem is communication obstacles. It is possible that I am giving some piecemeal information. Then there may be problems with his record. He doesn't know what to record, and may need further communication.} The same challenge also reported from another participant \quoted{G1P1-A: It will cause the lack of context.} and \quoted{G6P12-A: He didn't let me check in the end. I think he wrote a lot of error.} And G5P9-A said \quoted{Because I prefer to have frequent sentence breaks in the process of speaking a sentence. I will say a long sentence, and there may be many sentence breaks in it. Maybe this is a problem of my personal language habits. But the other party may just break it, and it may be a little different from what I said.}}

\paragraph{Authors' Wish-list.} We asked the Authors what they wished the Typists could have done. Most participants mentioned they hoped their Typists help them check and fix the grammar and typos. Three groups mentioned rather than recording silently, they hoped the Typists could continuously give feedback while they compose.
\added{As G5P9-A mentioned \quoted{Besides, I don’t know if he can keep up with me, I don’t know where it stops, and then I just stop for a while. I hope he will give me feedback after he finishes typing.}} More requests on the wish-list include: Typists marking their mistakes but not fixing them; checking the format and structure of their text; learning their preferences and habits in order to help them fill missing words. In addition, an Author preferred the Typists to not modify any of their dictation but mark them out. 
\added{G5P10-A explained that \quoted{For example, punctuation or grammar is to make this paragraph of my content more complete, but changing words is to modify the description and change my original intention. This may not be acceptable to me. I can accept that you help me fixing grammar and punctuation, but only if my original intention is accurately expressed and paraphrased.}}

\subsubsection{\newcl{Effects of modality for Authors (RQ4)}}
To investigate the impact of \modality{} on Authors' operations, we ran a T-test on each category of Author utterances between the \ef{} (AO) and \ws{} (AS) condition. To remind, the categories are: Create new content, Re-speaking, Explicit editing, Explicit navigation, Content review, Ask questions and Delegate Task. The results showed significant differences on \sy{three} categories: Explicit editing, Content review and Ask questions. No significant difference was identified in other categories. We will elaborate in the respective sections. We also performed a T-test between \ef{} and \ws{} condition for each type of Re-speaking, as identified in Section~\ref{sec:respeaking}, and identified no significant difference. In terms of subjective preference, the participants were asked for their preferred condition between \ef{} and \ws{} as Authors. \sy{Nine} Authors preferred the \ws{} condition while \sy{five} preferred \ef{}. \sy{Six} Authors had no clear preference.

\begin{figure}
\centering
\begin{minipage}[t]{.30\textwidth}
  \centering
  \includegraphics[width=0.7\linewidth]{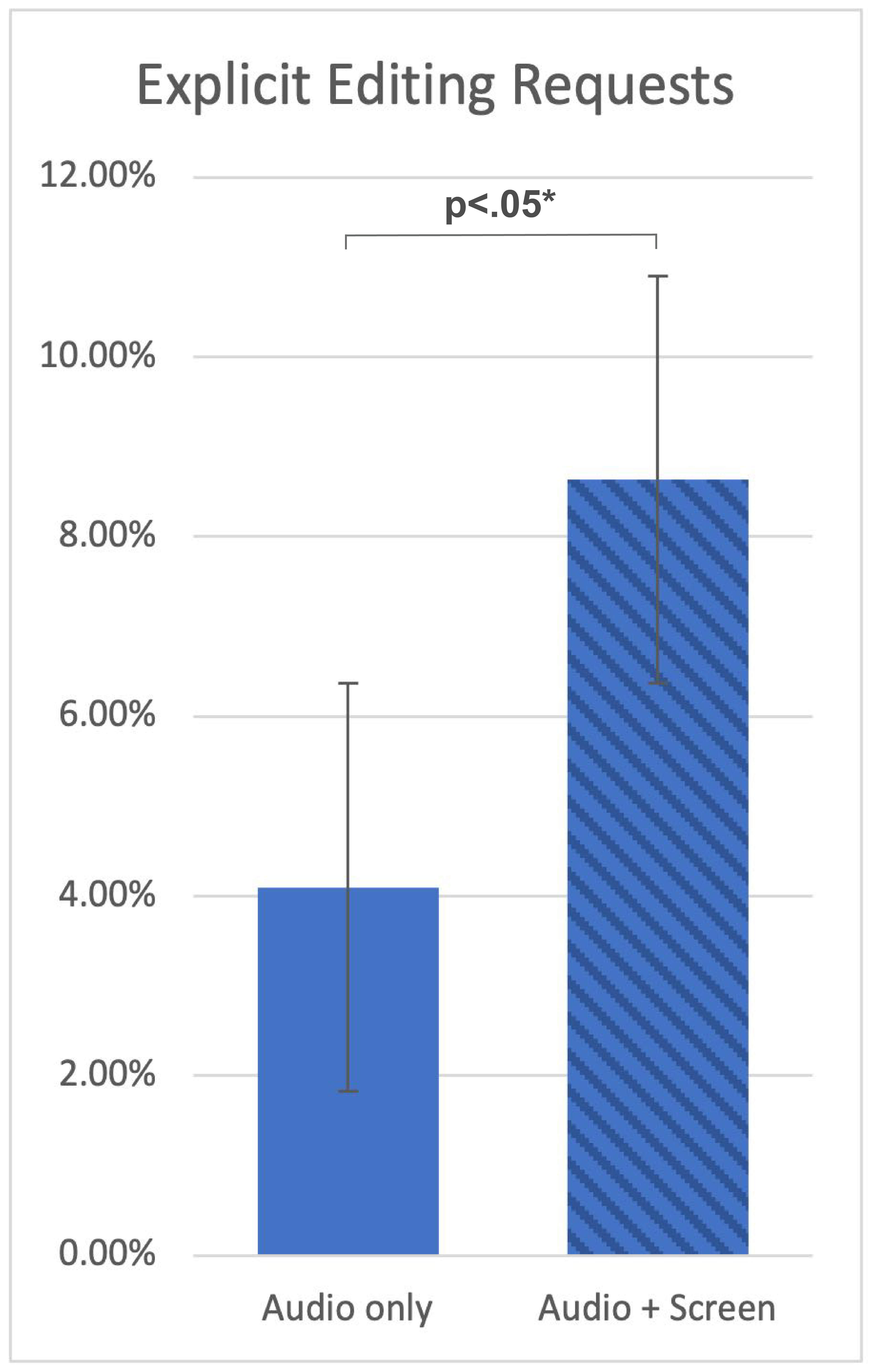}
  \captionof{figure}{Percentages of occurrences of Explicit Editing Requests in \ef{} and \ws{}.} 
  \label{fig:ef_ws_editing}
\end{minipage}%
\hfill 
\begin{minipage}[t]{.30\textwidth}
  \centering
  \includegraphics[width=0.7\linewidth]{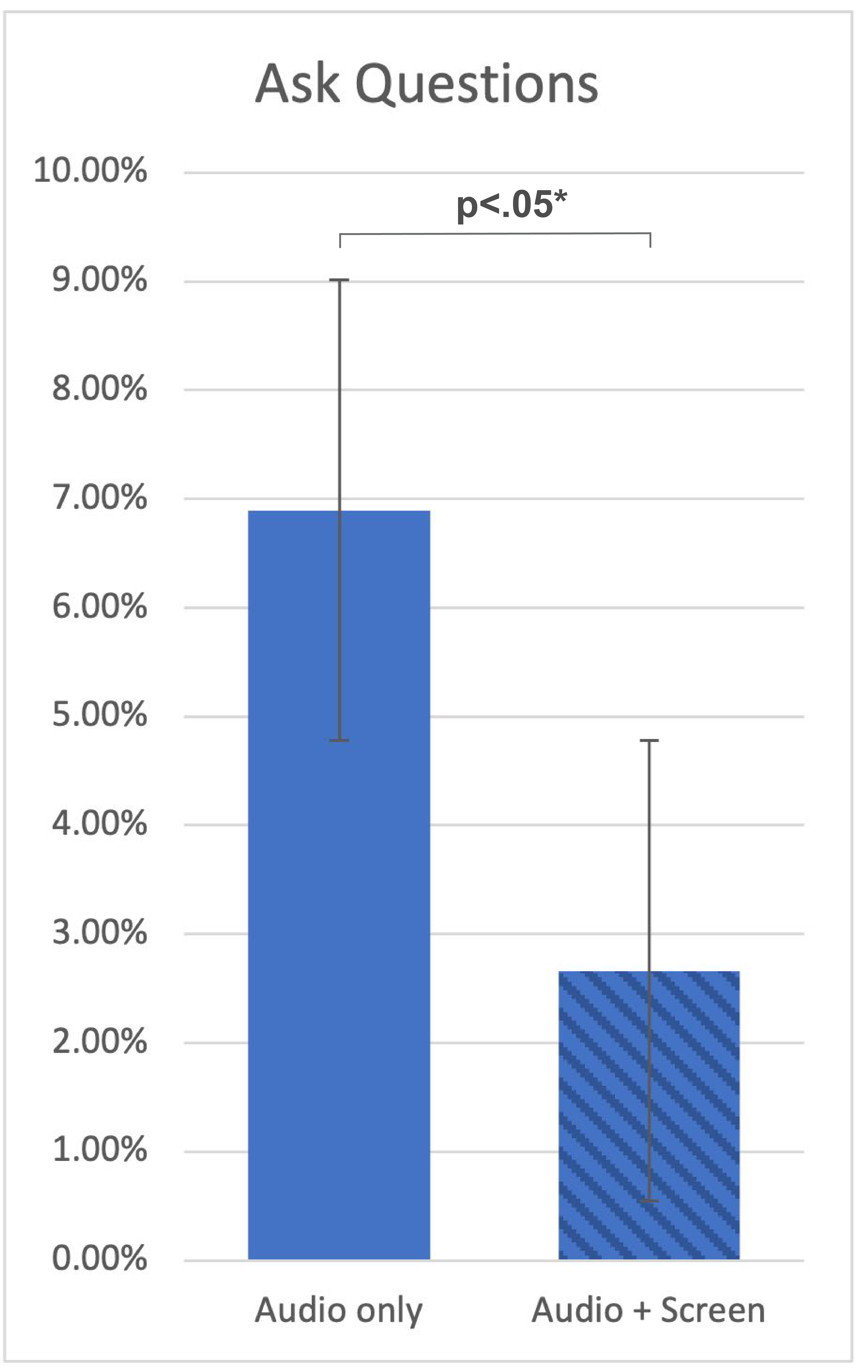}
  \captionof{figure}{Percentages of occurrences of Ask Questions in \ef{} and \ws{}.}
  \label{fig:ef_ws_askquestion}
\end{minipage}%
\hfill 
\begin{minipage}[t]{.30\textwidth}
  \centering
  \includegraphics[width=0.7\linewidth]{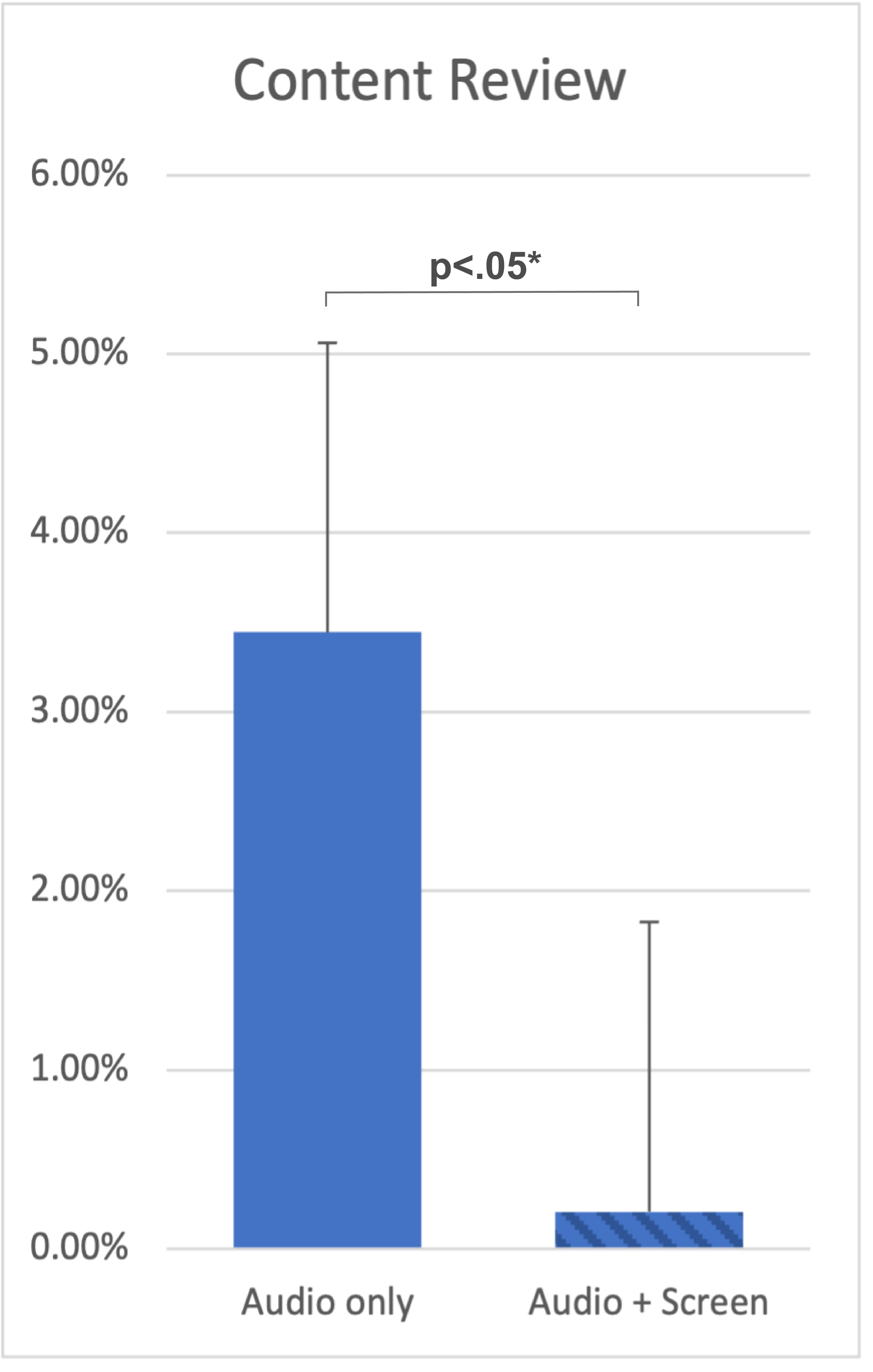}
  \captionof{figure}{Percentages of occurrences of Content Review in \ef{} and \ws{}.}
  \label{fig:ef_ws_review}
\end{minipage}%
\end{figure}

\newcl{\paragraph{Statistically significant differences.} Fig.~\ref{fig:ef_ws_editing} shows Explicit Editing were more frequently requested in \ws{} than \ef{}. A two-sample t-test was performed to compare the occurrences of Explicit editing requests showed a significant difference between \ef{} (M = 4.1\%, SD = 4.7\%) and \ws{} (M = 8.6\%, SD = 8.0\%); t(19) = 2.1, p = 0.012.  
Fig.~\ref{fig:ef_ws_review} shows significantly more content review requests occurred in \ef{} condition compared to \ws{}. A two-sample t-test performed on the occurances of Content review in \ef{} and \ws{} showed a significant difference between \ef{} (M = 3.3\%, SD = 5.6\%) and \ws{} (M = 2.7\%, SD = 1.0\%), t(19) = 2.1, p = 0.035.
Fig.~\ref{fig:ef_ws_askquestion} showed questions were more frequently asked in \ef{} than \ws{} condition. A two-sample t-test was performed on the occurances of Ask questions in \ef{} and \ws{} showed a significant difference between \ef{} (M = 6.9\%, SD = 6.0\%) and \ws{} (M = 0.3\%, SD = 3.2\%); t(19) = 2.1, p = 0.002.}

\paragraph{\newcl{Not seeing text lets the thoughts flow.}} Participants explained their reasons of preferring \ef{}. Two Authors said it allowed them to focus and keep their train of thought while composing, whereas looking at the text could be distracting. G5P9-A-AS articulated well: \quoted{When I can see him typing, sometimes my train of thoughts get interrupted, because I had to watch where he was segmenting sentences and where he put punctuation, I needed to make space in my head to think about those, it was distracting. When I cannot see, I just let him handle it and trust him.} G9P18-A-AS said something similar,\quoted{because it allowed me to focus more on the image and, composing the text in my head. I knew that [Typist name] could see what I was writing. So I didn't have to think about the cues and things like that. So it was in a way more relaxed.} In addition,  Authors mentioned it was less stressful and they had less self-doubt in the \ef{} condition. One Author said it felt more free, \quoted{G5P9-A-AO: I like not to see it. It feels more free, I don't get constrained.} Interestingly, the modality also affected the trust between Authors and Typists. \quoted{G3P5-A-AS: With the screen, I rely less on [Typist name]. But with the eyes free, then it's more collaboration. More trust in him.}

\paragraph{\newcl{Seeing text supports and encourages editing.}} Authors explained that having a shared screen helped them to keep track and make sure there was no misunderstanding. \quoted{G2P3-A-AS: I wish to see, because this is an interactive process, when I input something, it can be checked in time, it's an interface.}  One participant preferred \ws{} because the screen ``helped him to think'' (G9P17-A-AS). Beyond these, seeing the spoken text also makes Authors more compelled to edit. This partially explains why Authors made more explicit edits in \ws{} condition (Fig.~\ref{fig:ef_ws_editing}). G3P6-A-AO explained he was more easily satisfied when he could not see the text, but he cared more about visual details when seeing it. \quoted{well the eyes free condition. It just hindered my desire to edit, made it easier for me to be satisfied because the editing process was a bit cumbersome. (...) I'm happy with it. I noticed that they said visual components to text when you're writing, you're not, you don't only care about how things sound, but also how things look. So it's not the same to leave a line in between paragraphs or to go to the next paragraph, or to make a line longer or shorter. And that dimension disappears, when we have somebody reading to you. So you don't care about that. But if you see the text, then you have these, these other axes that you care about, and it's something that you're going to incorporate in the text edition.}

\subsection{How the Typists assisted the Authors? (RQ2)}

\subsubsection{\newcl{Categories of passive and proactive assistance}}

\begin{figure}[]
  \centering
  \includegraphics[width=\linewidth]{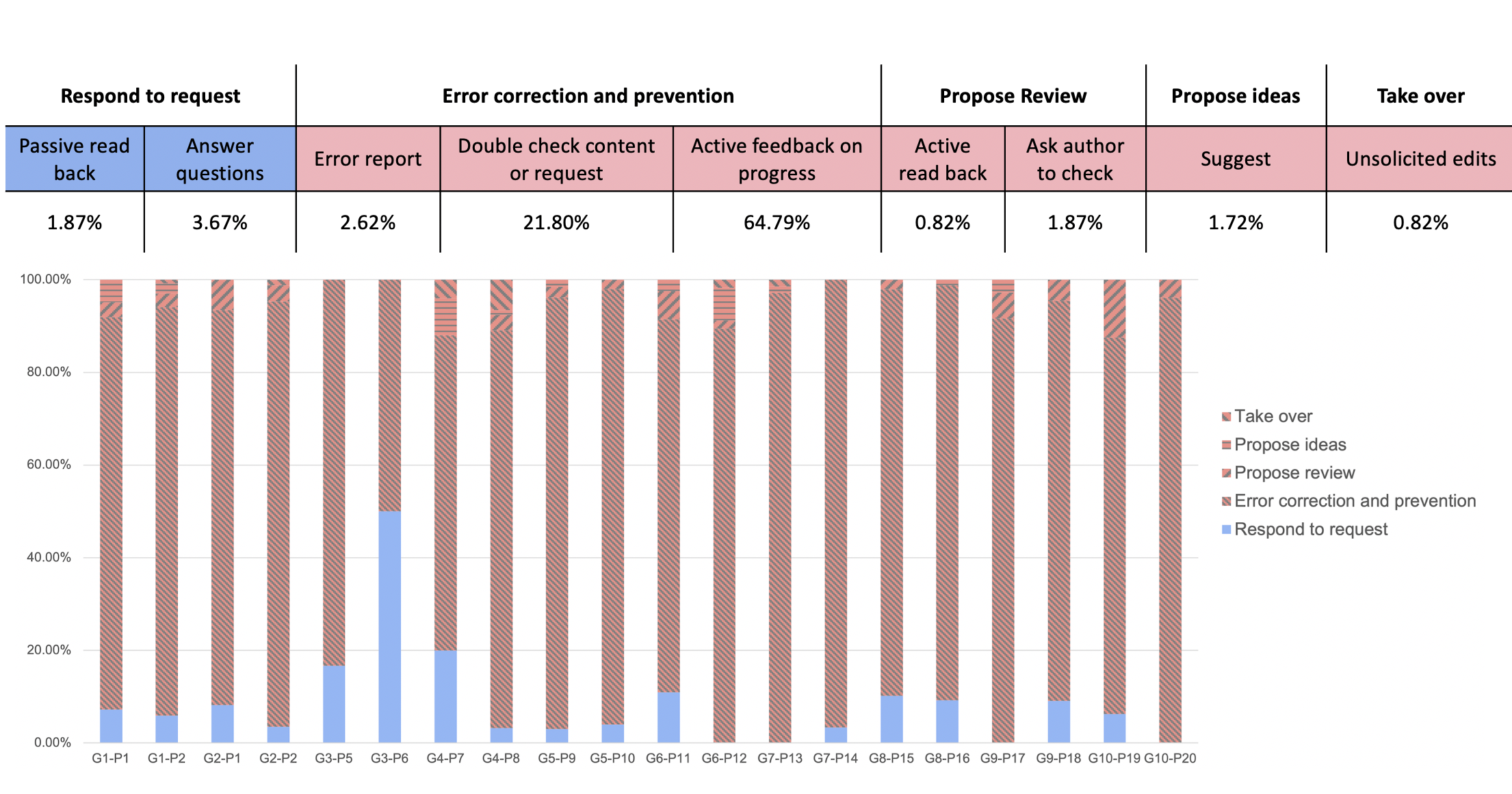}
  \caption{\newcl{Top: The categories of verbal assistance provided by the Typists and their occurrences in percentages. 
  Blue categories were passive responses. Pink categories were actively initiated by the Typists. Bottom: Visualization of the style of each Typist by stacking the percentage of each category of their utterances.}
  \Description{A spectrum of Typist behaviors ranging from passive to proactive. }}
  \label{fig:typist_behavior}
\end{figure}

Fig.~\ref{fig:typist_behavior} shows all the categories of Typist behaviors emerged from our analysis. \old{We observed a spectrum of behaviors ranging from passive to proactive.} Based on our observation, we summarized a higher level of categories to represent the types of ``services'' provided by the Typists, including: Respond to request, Error correction, Error prevention, Propose review, Propose ideas and Take over.
\newcl{While \textit{Respond to request} was observed to be passive, all the other categories describe active feedback or assistance initiated by Typists. As we can see from Fig.10, active assistance accounted for the majority of Typist utterances, leaving passive verbal responses only 5.5\% in a sum. Out of the active assistance, \textit{Error correction and prevention} was the most frequently observed behavior category, within which providing “Active feedback on progress” accounted for 64.8\% of the total utterances. We elaborate on each category in the next subsection. Furthermore, we observed various group dynamics, which affects the passive/active level of the assistance. Fig.~\ref{fig:typist_behavior} illustrates the stacked percentages of utterances in categories from each Typist. While most Typists were active in their verbal assistance, we can see G3P6 was the most passive, with half of the time only responding to the Author's requests and never made suggestions.

}

\subsubsection{``Services'' provided by the Typists}
The Typists demonstrated the following behaviors when assisting the Authors with the task. 

\paragraph{Respond to request}
Two types of passive behaviors of the Typists emerged. One is \emph{Answer questions}: Typists directly answer Authors' questions. The types of questions being asked are elaborated in Section~\ref{sec:author_ask_ques} \modif{and Appendix~\ref{tab:typist_codeutterances}}. The other one is \emph{Passive read back}, which happened when Authors asked Typists to read back part of the transcribed text for reviewing it. 

\paragraph{Error correction and prevention}
\label{sec:typist_feedback}
The Typists demonstrated \old{three} \sy{three} types of behaviors to help with error correction or prevention. \newcl{It included \emph{Error report} behaviors where Typists brought it up when something seemed wrong, \emph{Double check content or request} when Typists were unsure, and \emph{Active feedback on progress} to help Authors keep track and synchronize. }

The Typists reported to the Authors when it was clear something went wrong in the process, coded as \emph{Error report}. The types of reported errors include: the Typist did not understand which words to type, e.g., \quoted{Which word? I didn't hear it clearly}, \quoted{What do you mean?}; the Typist could not catch the speed of the Author, e.g., \quoted{I couldn't catch up!} \quoted{I forgot what's after ...}; the Typist did not know how to spell the word, e.g., \quoted{... I forgot how to spell it}.

Interestingly, the Typists also exhibited various types of Re-speaking behaviors for error correction or prevention. They were not re-speaking what they said by themselves, but what the Authors said. The Typists repeated some part of the content in a question when the Typists missed or misheard the words, e.g., \quoted{G10P19-T: `Very' what, sir?} More implicitly, when the Typists misheard some words in the end of a sentence, they simply repeated the last correctly-taken word using a question tone, instead of explicitly saying they misheard something. Take a look at this example. \quoted{G2P3-A: I would name the picture as a cat and a hard crystal ball. G2P4-T: A cat and a hard?} In this case it was clear the Typist missed the last few words. Not only for correcting mistakes, Re-speaking were also used to prevent errors when something did not sound right. The Typists double-checked content by re-speaking the problematic words with a question tone, e.g., \quoted{G1P2-A: she is... she’s trashing. G1P1-T: trashing?} 

In order to prevent errors from occurring, the Typists did \emph{Double check content or request}, following an instruction given by the Authors that appeared unclear or incorrect to the Typists. When it was about \emph{content}, besides Re-speaking, they asked questions like,\quoted{G10P20-T: Maybe somewhere `to' the future or `in' the future?} \quoted{G10P20-T: Did you say `it seemed to lead down' or ...? G10P19-A: Oh, `lead him'.} They also asked questions based on the grammatically structure, e.g., \quoted{G1P1-T: What was the subject of this sentence? `at the bottom of the books'.} Sometimes they simply asked the Authors to repeat the sentence they just said, e.g., \quoted{G1P1-T: Please repeat this sentence.} When it was about the \emph{request}, what was unclear could be about the \emph{location}, e.g., \quoted{G1P1-A: em, at the second sentence, next to `the soldiers'. G1P2-T: The second?} \quoted{G6P11-A: not `wine', I meant `wire'. G6P12-T: In which line?} The confusion could also be about the \emph{editing operation}, e.g., \quoted{G6P11-A: Continue deleting! G6P12-T: You don't want the following anymore?} Sometimes it became unclear whether it was about the content or the editing operation: \quoted{G6P11-A: Change the earlier thing to be `a boy is walking in the beautiful universe'. G6P12-T: beautiful universe, do I remove the `grass'?}

The Typists verbally provided \emph{Active feedback} to the Authors to help them keep track of their typing progress. There are \sy{three} types of them. The first one observed was that the Typists constantly read while typing, in order to provide real-time feedback of where he/she was. The second verbal cue was a brief signal indicating typing just finished, such as ``em''. Similarly the third verbal cue was phrased as a suggestion to continue, ``ok, and then?''. Last but not least, since we set a word limit for each task, the Typists sometimes reported the word count when they saw appropriate. This was rather necessary in the \ef{} condition.

\paragraph{Propose Review}
There were times when the Typists initiated a text review without being asked by the Authors. One way of doing this was \emph{Active read back}, which was the Typists proactively reading back the entire or a large part of the content without being asked by the Authors. G5P9-T explained for some unsure things, like the tense, he would read back instead of asking questions, \quoted{When it comes to tense, I wouldn't ask him which tense, I would read it back to him. If he thought it wasn't right he would correct it.} G1P1 also said he/she would read it back instead to double-check.

The other way of proposing a text review was to ask the Authors to check the content, coded as \emph{Ask \sy{Author} to check}. In the \ws{} condition, it was straightforward that the Authors were asked to look at the text, e.g., \quoted{G6P11-T: Please check if there is anything wrong.} In the \ef{} condition, the Typists had to read back the text to be checked, e.g., \quoted{G6P12-T: Let me read it back to you and see how you feel...} 

\paragraph{Make suggestions}

The Typists provided active \emph{Suggestions} to the Authors to help with the task. Content suggestions included grammatical corrections, e.g., \quoted{G5P9-T: I think the `and' should be changed to `which', right?}, better wording, e.g., \quoted{G6P12-T: How about I change it to `with different size'?}, or about the style, e.g., \quoted{G6P11-T: Another `and'? Let's not use `and' again, too many `and'!} Besides suggestions about the content, the Typists also gave procedural comments, for instance, \quoted{G4P8-T: Or let's edit this first, fix the capitalization and then come back to think about the last sentence.}

\paragraph{Unsolicited edits}
\label{sec:unsolicited}
\emph{Unsolicited editing} is the behavior of the Typists editing the text without consulting the Author. This happened mostly for correcting obvious mistakes or the mistakes of the Typist him/herself, occasionally for the meaning of the sentences. When being asked whether they did things the Authors didn't ask for, 8 out of 20 participants reported that they took the initiative and corrected minor grammatical errors, such as the use of singular and plural, prepositional and conjunctions, etc. \sy{four} of them took the initiative to revise the rationality or meaning of the sentence. 

In the interview, participants elaborated why and how they made unsolicited edits. For instance, G9P17-T corrected \emph{segmentation of sentences and captions}, \quoted{... I was probably having a little bit more ... control over say things like where the sentences ended and full stops because like [AuthorName] wasn't able to see that.} G5P9-T explained in detail her mental process while making segmentation: \quoted{I couldn't be sure whether he finished the sentence, so I would segment it subconsciously (with period). But then I felt the sentence after seemed to connect to the previous one, I would change the period to comma. I would follow my own judgement and I didn't ask him.} G5P10-T \emph{corrected his own mistake} once realized, \quoted{I heard `honor', then I felt something was wrong, ..., then I realized I spelt it wrong, so I corrected it later. At that time, [AuthorName] could not see.} The same participant also mentioned him adding \emph{plural}, \quoted{... I added `s' when I felt there should be one... But sometimes I would ask her when I'm not sure.} In addition, the Typists added \emph{punctuation}. \quoted{G7P13-T: Because I am not only able to listen, but also to think, to understand its content, and then make a judgement whether it should be a question mark or a comma.}

Some Typists refrained from making unsolicited changes, or even content suggestions. G9P18-T said, \quoted{I tried to be faithful to the \sy{Author}. So I tried not to add it. I was only edited in my own errors.} G3P6-T said, \quoted{the only thing I did was focusing on, on writing and being accurate. And then adding the commas and full stops wherever I felt they should be... because [AuthorName] didn't provide any information for that. ... I refrain from making any comments or suggestions about the text because it would have been inappropriate.} Interestingly, G3P5-T described her \emph{mental model} and experience of being a Typist as ``killing part of the brain'': \quoted{I used to have to take minutes as a teaching assistant or for work. So it's like, you kill part of your brain, like you stop thinking and you just type. But ... the problem is that sometimes [AuthorName] would say something and my head would say, but, like grammar wise that doesn't make sense. Or like, sometimes the grammar tried to override the thinking, the `not thinking'. It's like, if you get so used to hearing the way certain things are said for many years. It's hard to not change things. It's like an instinct.}

\subsubsection{Typists' choice of modality (RQ4)}
As Typists, \sy{eleven} participants preferred the \ws{} condition while \sy{four} preferred \ef{}. The rest \sy{six} of them had no preference. The main reason for preferring the screen was to get real-time feedback from the \sy{Authors} to be more efficient. \quoted{G2P4-T-AS: I felt a sense of safety, feel like what I wrote can receive realtime feedback. Sometimes there was mistyping, maybe that word is pronounced the same but spelled differently. If he can see it he can tell me.} \quoted{G5P10-T-AO: I hope he could see me typing, because it could avoid many problems, like homonyms or punctuation.} \quoted{G9P18-T-AS: So the screen was helpful that I could be sure to know that I was doing what she wanted me to do.} G1P1 thought it highly depended on the tasks: if she was recording exactly the Author's words, then \ws{} was preferred. If she needed to be creative in editing the content, \ef{} was preferred. Furthermore, \sy{two} groups of participants thought being able to see made it easier to locate text. 
As the reason to prefer \ef{}, G2P4-T-AS reported having mixed feelings about \ws{} because she felt uncomfortable and nervous that the Author could see her typing mistakes immediately; yet she found it effective for completing the task.

\newcl{
\subsubsection{Typists' perceived challenges}

The Typists also faced challenges in the process of assisting Authors. Four main challenges were identified in the interview. (1) Mishearing words or wrong spellings. As G2P4-T said, \quoted{Listening is also difficult, there are a lot of words that are pronounced similarly and then the network is not particularly clear, so it's easy to get mixed up and not know what words are there anyway.} (2) Misjudging punctuation due to a lack of understanding of the content being expressed. As mentioned by G5P10-T,\quoted{After reading the sentence, as a \sy{Typist}, I thought he had finished the sentence, but in the context of [AuthorName], she hadn't finished the sentence, so she decided to go on. But at this place I typed a full stop, but he was probably prepared to put a comma.} (3) Unsure about how to provide feedback. For instance G10-P2 was worried about providing constant feedback being annoying. 
\sy{(4)} Losing track of the \sy{Author}. This was sometimes due to the typing speed slower than the Authors' speech. Some other times it was due to Typists' memory, as G4P8 said, ``I might forget what the second half of what he said was when I was typing. Just asked him to say the second half of the sentence.'' }

\section{Between Authors and Typists (RQ3)}
The interaction between Authors and Typists demonstrated rich behavior patterns in terms of how they cooperatively navigated and located text, how misunderstandings occurred and were resolved, how they coordinated their closely-coupled collaboration and how they adapted to each other's work style.

\subsection{Coordination}

This dictation task is a closely-coupled collaborative task~\cite{wang2017users, posner1993people, fussell2000coordination}, 
which requires the Authors to keep track of the Typists and the Typists needed to be able to follow the Authors all the time. While this is relatively easy when visual feedback is provided, it can be a challenge in the \ef{} condition. 

\subsubsection{Navigating and locating text}
\label{sec:between_navigation}

When Authors asked Typists to edit, read back or address a question about a part of the text, they needed to specify a deictic reference to help locate the target. We identified the following methods they used for locating text. 

\paragraph{Deictic References.} Calling \emph{keywords} or reading a phrase was the most frequently observed way to locate content. If the keywords are not exactly the target, the reference is commonly accompanied by a \emph{temporal deixis}, namely ``before'' or ``after''. More observed temporal deixes include ``the sentence/words I just said'', ``at the beginning'' or ``at the end''. Actually, most of these temporal deixes used by our participants can be seen as \emph{spatial deixes} as well, given the sequential nature of textual content. \emph{Numeric references} were used in combination with a text unit, such as, ``the second sentence'', ``the last word'', ``the first line''. Locating with the order of \emph{lines} was only observed in \ws{} conditions. 
Example 1 below shows a few occurrences of temporal and spatial deixes. More examples 
can be found in \modif{Appendix~\ref{tab:author_codeutterances}}. 

\paragraph{Communicating with mouse pointer.} 
\label{sec:out_of_sync}
In \ws{} condition, apart from the  text and the moving cursor at the editing position provided visual feedback, we also noticed one more visual cue - the \emph{mouse pointer of the Typist}, played an important role in the communication. Example 4 below 
illustrates how the moving pointer of the Typist reflected how she was lost in searching for a keyword, where she was searching, and how the Author guided her smoothly from locating the sentence to the part of the sentence in subtle steps. Example 5 shows a situation when such visual feedback is missing and how it can become difficult to resolve misunderstanding. The Typist got lost at some point and asked a few questions to seek for help locating the editing position with keywords and numeric indexes of the sentence: \quoted{Do you mean after the sentence with `white socks'? Or `walking up a stair?' After the second sentence? Or after the first sentence?} But the Author could not directly answer these questions, probably because she could not see which sentence that was, thus continued to repeat the keyword `black hole' to locate. That did not solve the problem - the Typist responded with something even more confusing. The Author then gave up the precision and said `whatever', until later simply started a new sentence.

\vspace{5mm}
\begin{multicols}{2}[\small\textbf{Selected example dialogues between Authors and Typists:}]
\vspace{-3mm}
\tiny
\begin{quote}
            \textbf{Example 1}. G6, \ef{}.\newline
            P11-A: \quoted{\hlight{Grass}}\newline
            P12-T: \quoted{Grass, okay.}\newline
            P11-A: \quoted{`And blow soul bubbles, it seems that', \hlight{to the end,} don't use `it seems that', it seems that, 
            \hlight{Put this sentence what I just said, add a little thing before this sentence, go back to its beginning first.} `The picture describes a dream of a child, it is so simple for a child to get happiness.'}\newline
            P12-T: \quoted{So simple for a child.}
\end{quote}

        \begin{quote}
            \textbf{Example 2}. G6, \ws{}. \newline
            P12-A: \quoted{Comma, appearance, they give out, full stop, yes. they give out that they knew how to weave scarves of the most beautiful colors, scarves of, of the most beautiful colors, and elaborate patterns, \hlight{ELABROATE, ELA}, yes,\hlight{BR} , Oh,\hlight{ BORATE} , no more.}\newline
            P11-T: \quoted{\hlight{You can check to see if there is anything wrong.} \newline
            P12-A: \quoted{Change the full stop of the `leaf' to a comma.}\newline
            P11-T: \quoted{What line?}\newline
            P12-A: \quoted{\hlight{Second line}. Capitalize the 's'.}}
        \end{quote} 
        
        \begin{quote}
            \textbf{Example 3}. G5, \ef{}. \newline
            P10-T: \quoted{OK. A toy sitting on a chair manipulated by someone we don't know. Does it sound familiar? It’s just like my life. When I was a little boy If was manipulated by my parents...}\newline
            P9-A: [interrupting by barging in] \quoted{\hlight{No, sorry. I have to correct. It seems that you didn’t change the words I said.} Just does it sound familiar and actually it’s my real life.}\newline
            P10-T: \quoted{OK OK. You want to change the third sentence into `actually'?} 
        \end{quote}
        
        \begin{quote}
            \textbf{Example 4}. G1, \ws{}. \newline
            P1-A: \quoted{Em, at the second sentence, near `the some soldiers'. [P2’s \hlight{pointer} starts to \hlight{travel rapidly} through the whole text.] Em..} \newline
            P2-T: \quoted{Second? [P2’s pointer is \hlight{scanning the second sentence}.]} \newline
            P1-A: \quoted{... From the second sentence, next sentence, add a sentence. [P2’s pointer \hlight{stops at the beginning} of the second sentence.] ... \hlight{After} the second sentence. [P2’s pointer finally \hlight{moved to the end} of the second sentence.] Em. } \newline
            P2-T: \quoted{Em.} \newline
            P1-A: \quoted{`they hold the guns'.} 
        \end{quote}
        
        \begin{quote}
            \textbf{Example 5}. G1, \ef{}. \newline
            P2-A: \quoted{`Above his'... wait, the `walking'.} \newline
            P1-T: \quoted{[Do you mean here] `He is walking up a stair to a black hole'?} \newline
            P2-A: \quoted{Then you say `above him at the black hole', `above' as on the top.} \newline
            P1-T: \quoted{Do you mean \hlight{after the sentence with `white socks'? Or `walking up a stair?' After the second sentence? Or after the first sentence?}} \newline
            P2-A: \quoted{\hlight{`black hole', after `black hole'.}} \newline
            P1-T: \quoted{`black socks'? After `white socks'? Ah black hole, okay.} \newline
            P2-A: \quoted{Either is fine, \hlight{whatever}.} \newline
            P1-T: \quoted{Ok. Then, `above the' ...?} \newline
            P2-A: \quoted{`above'... \hlight{How about this, after the `black hole', after that, start a new sentence:} `Above him, him and the black hole,'} \newline
            P1-T: \quoted{`and the black hole'} \newline
            P2-A: \quoted{`comma, there are two'} \newline
            ... [P1 continues repeating after typing and P2 continues composing] 
        \end{quote}
        
        \begin{quote}
            \textbf{Example 6}. G2, \ef{}. \newline
            P3-T: \quoted{Glass. OK, I need to read again: `It tells the power of time to change the cute young girl to...'} \newline
            P4-A: [interrupting by barging in]\quoted{`Tells the power of time', \hlight{remove after that}.} \newline
            P3-T: \quoted{Sorry?} \newline
            P4-A: \quoted{\hlight{The last sentence.}} \newline
            P3-T: \quoted{\hlight{The last sentence:} `It tells the power of time to change the cute young girl to an old lady.' \hlight{Just delete, right?}} \newline
            P4-A: \quoted{Uh, \hlight{`it tells the power of time.'}} \newline
            P3-T: \quoted{OK. `It tells the power of time', \hlight{nothing}.} 
        \end{quote}
        
        \begin{quote}
            \textbf{Example 7}. G1, \ef{}. \newline
            P1-A: \quoted{`Hold by his hand. And \hlight{since he is walking to the mountain, since he is climbing the mountain.'}} \newline
            P2-T: \quoted{\hlight{`walking to the mountain', then, `climbing ... ' right?} } \newline
            P1-A: \quoted{Em... \hlight{`Since he is climbing the mountain.'}} \newline
            P2-T: \quoted{oh, oh, `since he is climbing the mountain', and then?} \newline
            P1-A: \quoted{And `some leaves is falling down from his body.'}

        \end{quote}
        
        \begin{quote}
            \textbf{Example 8}. G8, \ef{}. \newline
            P16-T: \quoted{OK, there is one sentence I didn't understand, \hlight{what do you mean by  `they look like very carefully'?}} \newline
            P15-A: \quoted{\hlight{They open their eyes extremely, very extremely, ... so big. Their eyes are so big.} [This was intended to overwrite the sentence asked by the Author.]} \newline
            P16-T: \quoted{\hlight{Which sentence exactly?} Which sentence? `Their eyes are so big.'} \newline
            P15-A: \quoted{Actually \hlight{just there,  at `they, they'} [he meant where the \sy{Author} was at].} \newline
            P16-T: \quoted{...[A few seconds of typing sound]... Alright?} \newline
            p16-T: \quoted{Alright.}
        \end{quote}
        
        \begin{quote}
            \textbf{Example 9}. G8, \ws{}. \newline
            P16-A: \quoted{`The direction of sunset. He is wearing a cloth made from leaves. Leaves.'} \newline
            P15-T: \quoted{Yeah.} \newline
            P16-A: \quoted{`In front of him are three', \hlight{no, this is the second sentence, a new sentence.} [Then the Typist added period and changed the `i' of `in front of' to `I'.]} \newline
            P15-A: \quoted{`three big mountains.'}

        \end{quote}
        \label{tab:example_dialogues}
    
\end{multicols}

\paragraph{Barging in readback} In \ef{} condition, the above-mentioned deictic references are not sufficient. \emph{Barging in readback} emerged as the Authors interrupted the Typists during his/her readback of the written text and suggested something near the interruption point. This was occasionally observed for locating problematic area, as seen in Example 3 and Example 6. 

\subsubsection{Synchronizing speed and matching rhythm}
In the Section~\ref{sec:typist_feedback} we described how the Typists provided \emph{active feedback} to inform Authors where they were. This behavior was indeed helpful according to our interviews. For instance, G9P17-T said, \quoted{I think just reading through at the end was useful for making sure that you got everything.} G4P7-A said, \quoted{She would read after every sentence. And she would read it again after finishing.} If the Typists did not automatically give active feedback, some Authors would ask for it, even defined their own verbal cue for it. \quoted{G4P8-A: I told him to make a `zhi' sound every time he finished typing.} Furthermore, Authors paid attention to the \emph{non-verbal cues}, namely the sound of the keyboard to make a judgement and adapt their speaking speed. A few more proactive Authors was \emph{actively asking} the Typists whether each keyword had been taken down, in order to track progress. In addition, there were Authors who \emph{did not pay active attention to track} the Typists' progress, relying on error calls by the Typists if they could not follow. G7P13-A simply slowed down her speech based on her own estimation: \quoted{I would compare to my own typing speed.} G10P19-A had full confidence for his Typist: \quoted{I mean [AuthorName]'s English level was phenomenal, and I feel that we were concentrated during the exercise.}

\paragraph{Typists' self-correction in idle time} 
We observed incidences where Typists reviewed the text spontaneously and corrected errors while the Authors were pausing to think or had completed. G9P17-T corrected segmentation of sentences and captions: \quoted{So thinking of something I might go back and correct where I put a full stop of sentence or correct the type while [AuthorName] was thinking. } G7P13-T explained the use of idle time to correct her own mistakes,\quoted{Oh, most of the mistakes I made when typing... [I remember them], and I would go back and correct it when I have a chance during task or when I have time.} 

\subsection{Confusion and Misunderstanding}
The participants generally felt their communication was smooth and misunderstandings were rare and only about small things. Yet, \sy{four} types of misunderstandings between Typists and Authors were observed during the experiment and reflected in the interviews.

\subsubsection{Mishearing words.} The primary cause of misunderstandings was Typists misheard words. They were particularly prone to this error when \old{\emph{synonyms}} \modif{\emph{homonyms}} were involved in the text or words were pronounced incorrectly. G10P19-T said, \quoted{I think there was also one moment where I didn't pronounce something correctly or came up with a name like `blob', so I felt like these were either slight misunderstandings or they were potential misunderstanding. So that's why I tried to correct them immediately and to be more clear for [AuthorName] to help him understand and bring some order to the chaos.} This quote also explained why \sy{Authors} repeated themselves as a measure for \emph{error prevention}. One participant explained this challenge was sometimes due to the \emph{lack of context}: \quoted{G1P1-T: When you (the Author) are talking, the \sy{Typist} doesn't know the context. If you are talking about a garden, then talking about flowers and trees is normal, he may know this word is flower or tree. Then if you suddenly say in the garden there is a Humanoid robot, maybe he couldn't associate immediately and just guessed it's another word more related to nature. So it's a problem of lost of context.} G3P5-T thought their team had \emph{minimal misunderstanding} and attributed that to their \emph{similar background and writing skill}. She said, \quoted{I think the misunderstandings ... when we pronounce certain words differently from each other. And to our idea of where to put comma, or period is different. I don't think there is any, like, major misunderstanding. And I know, this is not done on purpose. But the thing is, in terms of our background, you know, [AuthorName] and I are kind of similar. You know, we both have PhDs, we both were in [university name], we both have to do a lot of writing. And we both have to do a lot of creative thinking. [...] I'm guessing he and I ended up reading a lot of the same kinds of things. So, you know, when he says something, it makes sense to me. And when I said something to him, yes, it made him laugh, but he seemed to understand why I said it, and how I said it.}

\subsubsection{Misunderstanding of editing location or scope.} The location of requested edits could be misunderstood, especially when Authors and Typists got out of sync. When Authors moved their attention to a different area, they sometimes expected the Typists to follow, which could be unrealistic. Misunderstandings in location also occurred when the Author gave an abstract location that covered multiple sentences. In Section~\ref{sec:out_of_sync} - Communicating with mouse pointer, we elaborated examples of such misunderstandings and how they got resolved in \ws{} and \ef{} conditions. Even when the Typist correctly located the editing position, misunderstandings of which exact words to edit still happened. Example 6 shows one case of confusion and clarification about the scope of the deletion. The Author wanted to delete half of a sentence, but the Typist thought the intention was about the whole sentence. The Author corrected it by re-speaking the final text:``it tells the power of time.'', which then needed to be double confirmed with a repetition plus ``nothing'' by the Typist.

\subsubsection{Composing or editing?} Misunderstandings also occurred when the Author switched between composition and editing. This happened more often to some Authors who preferred to re-speak for overwriting content instead of giving an explicit editing request. Re-speaking utterances could be easily misunderstood as a new composition. For instance, Example 7 shows how an implicit editing of replacing ``walking to'' with ``climbing'' was misunderstood as new composition. Example 8 shows an example of even more confusing behaviors from Authors. When the Typist posted a question about a problematic sentence, the Author did not explicate anything, but directly spoke a new sentence to replace it. To make it worse, in the same utterance of speaking the ``replacement sentence'', he overwrote part of it three times and repeated the sentence once. Yet, the confusion was resolved surprisingly easily, he simply answered with keywords to help the Typist locate the editing position, and the Typist then understood everything. Here it must be due to the semantic similarity of the erroneous sentence and the replacement sentence. 

\subsubsection{Unclear sentence segmentation.} The last source of misunderstanding was reflected on Authors and Typists having different ideas about sentence segmentation. Example 9 shows one case of Authors fixing the sentence segmentation when they could see the text. The ambiguity of sentence segmentation was also reported in the interview, G10P19-T said, \quoted{I guess my main misunderstanding was just I think the sentence was ending or it was just like pausing.} G10P20-A explained this could also be caused by Authors being indecisive about where to end the sentence: \quoted{I guess when I was speaking, I kind of sometimes started a sentence and I realized I didn't wanna continue. So maybe I was a little indecisive when I was trying to tell some of the stories.} The misunderstanding of sentence segmentation appeared rather acceptable considering most participants thought their communication was smooth. G8P16-T said, \quoted{This does not affect understanding of the story, unless there are a couple of obscure words we couldn't understand. (...) There wasn't any big problem, just that we may wish the format looked the same as we imagined, (...), but it's just a matter of one sentence versus two.}

\subsection{Co-adaptation}
The cooperation strategies between Authors and Typists developed over time as they got familiar with each other throughout the experiment. The participants also applied strategies after they switched roles to better facilitate the partner, as they experienced some challenges on the other role.  

\subsubsection{Adjusting their speed and speech} The co-adaptation between Authors and Typists was first reflected on how they adjusted their speed to fit each other. \sy{Three} groups mentioned in the interview that they slowed down the dictation speed as an Author and became more patient. One participant reported he started to pay more attention to speak out punctuation after being a Typist. Some Authors started out dictating faster than what the Typists could follow, and slowed down over time to adapt to the typing speed. G7P13-A mentioned this was easier with the shared screen: \quoted{If [I] can see it, I wouldn't need to speak very slowly, I could automatically adjust and match my speed.}

\subsubsection{Adjusting feedback} As we explained earlier, both Authors and Typists needed to consciously provide feedback to each other. Authors needed to be informed of the typing progress in the \ef{} condition. Typists wanted effective feedback from the Authors when they made typing mistakes or misunderstood the dictation. A few Typists started out by passively providing feedback while being frequently asked by the Authors ``Have you finished (typing)?'', then gradually became active in reporting it without being asked. One participant explained how he adapted to the Author over time by reducing disturbing behaviors and refraining from making suggestions as well as getting more comfortable with typing. G10P19-T: \quoted{At the beginning, when I was at this automatically, I started kind of reciting after [AuthorName] so I was telling him what I was doing, which later on, I thought that's pretty annoying, but he should be in his own kind of creative space. Towards the edge, I felt that I needed to shut up, not to give suggestions. Plus I got used to the voice (...), his way of kind of coming up with the words. I feel more comfortable with all this thing to what do you want to say in terms of me being able to get it on paper as well.}

\subsubsection{Learning habits and preferences} The participants learned their partner's \emph{preferences and wording} over time, such as their vocabulary, whether they wanted more or less feedback, etc. As described in Section~\ref{sec:unsolicited}, Typists sometimes performed unsolicited actions and changed the text without consulting the Authors. This happened more in the later stages of the experiments, when participants were confident that such unsolicited actions would be accepted. G10P20-T:\quoted{ at the 4th story, I felt like I could maybe create a little bit more.} Verbal coordination also reduced over time, as participants began to understand what their partner meant or wanted without a thorough explanation. In the interview, one participant mentioned that getting used to her partner's vocabulary and wording habit would help reduce misunderstanding. \quoted{G9P17-T: (...) it's probably just a matter of getting used to someone's vocabulary. And then, if I listen to \sy{[AuthorName]} speaking more (...) if we did this long term, I think there would probably be less misunderstandings. So just getting used to what sort of words you tend to use for things. I might be probably better at guessing. }

\newcl{
\subsubsection{Influences of the modality}
Regarding how the Authors and Typists collaborated, modality first influenced how they helped each other navigate and locate text. In the \ef{} condition, the Authors either barged in the readback from Typists or relied on vague temporal / spatial memory about their composed text. In \ws{} they could locate precisely with keywords and other positional references. The Typists' moving mouse pointers were used as an effective communication tool for resolving misunderstandings in location and editing scope. Second, modality affected how Authors could keep track of the progress of the Typists. When Authors could not see the text, constant verbal feedback was needed from the Typists side. Having the screen also made it easier for the Authors to notice mistakes made by the Typists and provide feedback. }

\section{Discussion}
In this section we summarize our findings in terms of how they answer our research questions, and discuss them in light of developing future systems. 

\subsection{RQ1: How do Authors dictate to Typists?} 
Our findings answer RQ1 with a detailed understanding of the Authors' intentions, expressions, strategies and challenges when completing a dictation task assisted by another human. Without implying necessity, they suggested a set of experiences that could be potentially supported in future dictation interfaces.

\subsubsection{Diving into the ambiguity}

Our analysis dived into the ambiguity of human speech for dictation and visualized how implicit and explicit instructions were made throughout the development process of the written text. This approach differs from the data collection and analysis of natural speech with the goal of training a machine for speech recognition and repair \cite{10.3115/981574.981581}. \newcl{We identified both explicit and implicit ways that the Authors instructed the Typists to create or edit text. Explicit editing included requesting \textit{Add, Delete, Replace, Format, Organize, Punctuate} operations via informal and free speech. Implicit editing was instructed via \textit{Re-speaking}, which was a highly ambiguous behavior with five possible intentions including \textit{editing}, \textit{confirming}, \textit{locating}, \textit{continuing composition}, and \textit{repairing speech}. Furthermore, re-speaking was also observed on the Typists’ side, but in the context of repeating  the Authors’ words, as an implicit expression for \textit{error correction or prevention}. Recent state-of-the-art techniques support re-speaking phrases for eyes-free text editing ~\cite{ghosh2020commanding,fan2021just} 
or in multimodal interfaces ~\cite{ghosh2020eyeditor,zhao2021voice}. 
While these make important steps towards supporting natural dictation, our research reveals there is much more to understand about re-speaking and contributes a detailed understanding of the demands and noises around it. More research is warranted to explore how to support implicit user instructions.} 

\newcl{Another source of ambiguity was found in distinguishing composition and editing. We observed how the Authors went back and forth between composition and editing, by \textit{globally making several passes of the text} and \textit{locally mixing text composition and editing}. Our visualizations (Fig.~\ref{fig:content_strategies} and Fig.~\ref{fig:content_viz}) illustrate how entangled these \sy{two} modes are in natural dictation.} It is even harder to clearly distinguish between composition and editing. For instance, Re-speaking as an implicit editing behavior to continue the half-sentence spoken before, could also be considered as an edit of that sentence with an addition at the end. Existing dictation systems handle text input and editing with \sy{two} different modes that need explicit switching by users, which apparently does not adapt well with natural dictation. \newcl{Our appendix and supplement material of this work provides details of how the Authors and Typists communicated and how each piece of text got composed and edited over a timeline. They could serve as assets for system designers, engineers and researchers for further investigation or design consideration.}

\subsubsection{Dictation is not only about transcription}
Besides composing and editing, we also identified other behavior patterns of the Authors supporting the task, including \textit{explicit navigation, content review, ask questions, delegate task and think aloud}. 
While we expected the frequencies of behaviors being affected by individual choices in a semi-controlled experiment, we believe the list provides new perspectives of Authors' needs for natural dictation. 
We provide categories of expressions with examples used by the Authors, which can serve as inspirations for the design of new dictation interfaces. For instance, a conversational interface could potentially be provided to help users review content and ask questions. Moreover, our interview revealed a number of dictation strategies the Authors developed, including consciously reducing uncertainty and repetition, formulating and organizing their thoughts before speaking to avoid major changes, speaking slowly in simple words, etc. These can be seen as areas where users could learn or be willing to be trained on. 

\subsubsection{Help Authors organize and remember thoughts}
We identified a number of challenges the Authors faced, even with the intelligent assistance provided by Typists. The primary one was how they felt rushed and disorganized when composing text with speech.
Written text is a linear representation of the often arbitrary human thoughts. Human speech is produced as we speak, not before \cite{levelt1999producing}. Some Authors reported the challenge of organizing their free speech into a linear story, and wished the Typists could help. One Author suggested that the Typist could note down the Author's multiple composition ideas and later check with the Author. Furthermore, as the composition process developed, it appeared to be hard for most Authors to remember their own wordings without seeing the composed text. They used keywords as reminders or developed a spatial memory of the sentence or part of the text their target might be in. These findings could serve as inspirations for future intelligent dictation system features.

\subsection{RQ2: How do Typists assist Authors?} 

\newcl{Our findings answer RQ2 by listing the types of assistance provided by the Typists and elaborating their intentions, expressions and encountered challenges.} 

\subsubsection{\newcl{Error correction and prevention are major}}
\newcl{The majority of Typist utterances (over 85\%) were for correcting or preventing errors. Among them,  about 65\% provided active feedback about their typing progress.}  
It appeared that the active verbal and non-verbal feedback from Typists was crucial for Authors to keep track of the composed texts and coordinate with Typists. Automatic reading back, which is a feature being supported by previous work in eyes-free dictation~\cite{10.1145/3173574.3173977}, was shown to be an effective measure for both error correction and prevention. 
Future interfaces should improve on the timing and choices for voice readback. \newcl{More than reading back, the Typists frequently double-checked with the Authors about their intention, the editing location and  their actual editing request. Re-speaking potentially erroneous words with a questioning tone seemed to be an effective expression for double-checking. This could be implemented as a system feature.} 

\subsubsection{Proactive AI features are promising} 
\newcl{Except one participant being relatively passive, we found all the Typists were proactive in more than 80\% of their utterances (Fig.~\ref{fig:typist_behavior}).
Specifically, their proactive behaviors included \textit{correcting and preventing errors}, \textit{proposing review}, \textit{proposing ideas} and \textit{taking over}. Except for taking over with unsolicited edits that might cause a sense of losing control, this proactive assistance was largely appreciated and desired by the Authors, especially when Authors could not see the text. Many of the corresponding system features are already supported by existing systems by highlighting wrong words or automatic correction, such as Grammarly\footnote{https://www.grammarly.com/}.
Moreover, Typists’ self-correction appeared to be a needed feature. This is comparable to the automatic correction features based on semantic context in existing speech recognition engines such as the Google Cloud API\footnote{https://cloud.google.com/speech-to-text}. In addition, we found it was not easy for the Typists to segment sentences as how the Authors wanted without explication, even with the ability of understanding context as humans. Future systems could focus on engaging users in explicating segmentations.}

\subsubsection{\newcl{Effective Typist assistance}}
\newcl{Participants mentioned that it was helpful when Typists provided active feedback to Authors to inform their progress. One group with low level of misunderstanding thought it was because of their similar background and writing skill. Some apprecated the help from the Typist in coming up with better wording and segmenting sentences. Furthermore, participants appreciated the experience getting smoother as they adapted to each other over time and needed less explicit coordination. 
Compared to machines, humans have superior empathetic skills to detect what the other person needs during their interaction. The information comes from not only the words they hear but also the tone, the context of the task and their understanding of the other person. An even more powerful skill is that humans adjust their behaviors quickly based on the other person's reaction. If a Typist tried to provide an assistance and noticed it was not helpful, he/she would stop it quickly. If some critical support was missing, the Author would have asked for it. We believe the intrinsic assistance provided by most Typists can serve as a good reference for effective support to Authors.}

\subsection{RQ3: How do Authors and Typists coordinate and collaborate?} 
Our findings answer RQ3 by analysing how the Authors and Typists kept track of each other, how they communicated their intentions and resolved misunderstandings, as well as how they adapted to each other.

\subsubsection{\newcl{Coordinating location and communicating intention}}
By observing how Authors and Typists coordinated their actions in this highly synchronous collaborative text composition task, we identified 3 important factors at play: \textit{how to locate and select text with deictic references}, \textit{in what scope the Authors were operating (word, phrase, clause or multi-clause)} and \textit{the attention locale of both parties}. 

\newcl{To locate text, both Authors and Typists referred to one-dimensional temporal or spatial references for locating text, together with calling keywords and naming numeric indexes of sentences. Verbal barging-in was observed for locating text when Authors could not see the text. We believe these text selection concepts are fundamentally different from how we search and locate text in Graphical User Interfaces (GUI). However, our dictation interfaces today look very much similar to a traditional text editor, which is a GUI legacy centered around the cursor position. However, speech is more about words, semantics and subtleties in voice. We need to rethink the fundamental concepts of an editing interface adapted to speech input.
}

\newcl{Problems occured when it was unclear where the typing or editing location should be, how much of the original text should be removed or replaced, or when the Author and the Typist were looking at a different place in the text. To coordinate, the Typists and Authors gave active feedback to each other to synchronize their speed of speaking and typing, sometimes even establishing their own verbal cues to constantly signal the completion of typing. One effective communication strategy we observed when the Author could see the text, was that the Typist used the mouse pointer to circle around where she was editing location. Future systems could adopt similar strategies to assist error recovery interactively. For example, gaze input can be considered in such scenarios to help detect users’ editing intention.} 

\subsubsection{Unavoidable misunderstandings} 
Resolving misunderstand is important for both human-human and human-computer communication. First, we learned the types of misunderstandings, which were \textit{mishearing words}, \textit{misunderstanding editing location and scope}, \textit{confusion between composing and editing} and \textit{sentence segmentation}. Considering humans' superior ability of understanding context compared to machines, we believe \newcl{misunderstandings are unavoidable, and} more research is needed to understand how humans resolve these misunderstandings, such as what and how they use subtle behavior signals, to inform the design of intelligent machine support.

\subsubsection{Co-adaptive features}
The cooperation between Authors and Typists were like dancing, where they adjust to each others' speed and rhythm in real-time. They also adjusted the frequencies and types of feedback to each other by learning their habits and preferences from the reactions. Because of this, the participants felt there were fewer misunderstandings and smoother experiences developed over time. Future dictation systems could consider learning such rhythmic dynamics from the users' choices of words, composition speed, and feedback.

\subsection{RQ4: How does the communication modality affect Authors, Typists and their cooperation?}

Answering RQ4, we found both quantitative and qualitative differences between \ef{} and \ws{} conditions. Authors made more explicit edits with a screen compared to audio only, but similar amounts of implicit edits via re-speaking. One explanation indicated in the interview was that Authors were more tolerant to errors when they did not see the composed text. We also found the Authors asked fewer questions and requested less content readback from the Typists when seeing text. Participants also felt \ws{} helped them to be more efficient in finding and correcting errors, while in \ef{} Typists had to provide more feedback about the progress. Coordinating locations were also easier with a shared screen. It allowed creative communication strategies to happen, such as used the moving pointer to resolve navigation issues. 
Interestingly, despite it being frustrating to not be able to see the composed text, some Authors preferred the \ef{} condition. This was because not seeing the text can be liberating and allowed their thoughts to develop more freely. Seeing the text all the time was considered to be distracting. In addition, we found modality was associated with emotion. Not seeing the composed text could be less stressful and even increase the Author's trust in the Typist. 

\newcl{Based on our findings, we believe dictating eyes-free and with-screen should be both supported in future interfaces. These \sy{two} situations have different benefits and drawbacks that complement each other. Composing text in an eyes-free modality can be liberating, yet being able to see the system status is efficient and much needed for error correction. Too much visual demand would interrupt the eyes-free experience and break down some mobile user scenarios. Therefore we believe, how to strike a balance between seeing and not seeing the text, and how to provide visual and acoustic feedback, are major design and research challenges for future dictation systems.}

\subsection{\newcl{In the context of writing support research}}

Previous research has supported writing tasks in various ways. Numerous works have developed AI systems for automatically generating scripts or stories \cite{riedl2010story}. Researchers have developed intelligent tools to support users in creative writing. For instance, ReQUEST~\cite{riedl2008toward} is an intelligent tool for authoring plots, not by providing content, but by acting like an audience to provide users with constructive feedback. The system asks ``why'' and ``Consequence'' questions to direct the users through the process. Furthermore, computational analysis has taken place in extracting writing strategies from alargeamount of articles~\cite{august2020writing}. New writing strategies such as microwriting has been studied in the context of gamified writing ~\cite{iqbal2018multitasking}, or to support collaborative writing~\cite{teevan2016supporting} and crowdsourced writing~\cite{nebeling2016wearwrite}.

\newcl{
Although our research supports text composition tasks, we need to clarify that our focus is on dictation for \textit{text input}, which is different from other existing research on writing support that focuses on \textit{creative writing}. However, our findings contribute to the understandings of using speech to write, including the entangled composition and revision process, the ambiguity and disorganized nature of speech, and the liberating experience of speaking without looking at the text. Future AI systems could build on these understandings to develop creativity support for dictation interfaces.} 

\newcl{
Our findings on coordination and communication between Authors and Typists align with those in the literature of collaborative writing. For instance, previous research showed that collaborative writing can benefit from tools providing conversational grounding (e.g., enhanced coordination and group awareness)~\cite{lowry2003using}. To enhance conversation grounding in collaborative writing, Kütt et al. developed a tool to visualize their partner's gaze information, as a form of shared visual information, and showed that it helped to improve mutual understanding and flow of communication~\cite{kutt2019eye}. }

\subsection{Limitation}
\subsubsection{English fluency}
Our study might be affected by the fact most participants are non-native English speakers and they performed English writing tasks. Some exchanges between the Authors and Typists were about correcting grammatical mistakes and spelling words, which appeared less in native speakers. However, our native speaker groups of participants exhibited similar behavior categories. We did not observe visible biases to our main findings. Future studies will investigate potential effects of native versus non-native language use. While English fluencies and writing skills may affect quantitative ratios of different types of Author requests or Typist behaviors, we are confident the categories of phenomena are not affected. We draw our focus on the types of communication behaviors, which are in a way orthogonal to the actual content of the communication.

\subsubsection{Ecological validity}
\newcl{
This study explores human-to-human dictation, where Typists typed much slower than a speech recognition engine, and had limitations in memory capacity and English vocabulary. These potentially affected the behaviors of Authors in a number of ways. For instance we observed the Authors slowing down their composition to wait for the Typists, and the Authors occasionally spelled the words to the Typists. The slower typing speed and limited memory of Typists probably affected the efficiency of completing tasks, the frequencies of the Authors repeating content for the Typists, and the report from Typists for not following the Authors. 
In addition, the lack of English vocabulary probably led to more questions from the Typists and spelling help from the Authors, and maybe choices of simpler wordings.}

\newcl{We acknowledge that humans interact with computers differently from interacting with each other. There are emotional and social interactions between humans, which may not transfer to human-computer interaction in the same way. In this study we coded the data focusing on the actions and information flow related to task productivity only, not other aspects of the interaction. By finding out what support features Authors found effective, we hope to 
inform and inspire the design of future dictation interfaces. However, the actual effectiveness of the potential features still needs to be tested in a human-to-computer setting.}

\section{Conclusion}
This paper presents an experiment that investigated natural human dictation for text composition by observing pairs of participants dictating text to each other in the roles of Typists and Authors. Our results unpacked the ambiguity of natural dictation behavior and showed that supporting dictation is not just about improving speech recognition accuracy. Based on a comprehensive understanding of how Authors dictated, how Typists assisted and how they collaborated with each other, this work informs the design of future dictation interfaces by uncovering design opportunities and providing inspiration from human-to-human interaction. We are the first to use a role-play method for this purpose by observing how people provide an ``intelligent service'' to each other. It provides new insights that complement existing findings from Wizard-of-Oz studies and opens up discussions for future research on natural human behavior. 

\begin{acks}
This research was supported by the Hong Kong Research Grants Council - ECS scheme under the project number CityU 21209419. We thank our reviewers for their constructive feedback. 
\end{acks}

\bibliographystyle{ACM-Reference-Format}
\bibliography{TypistComp}

\appendix

\section{Codes for Author utterances}
\label{tab:author_codeutterances}
\begin{figure}[h]
  \centering
  \includegraphics[width=0.95\linewidth]{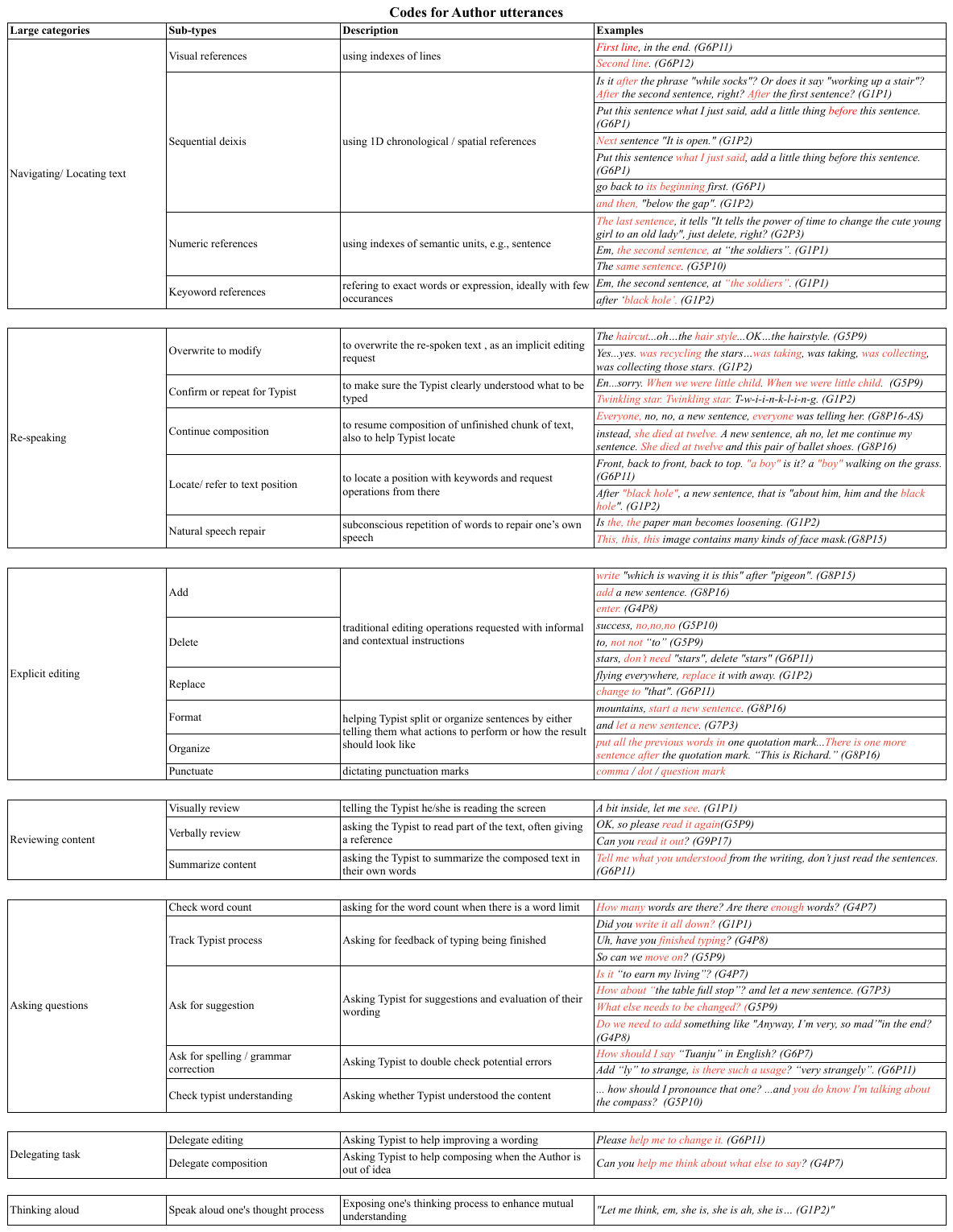}
\end{figure}

\section{Codes for Typist utterances}
\label{tab:typist_codeutterances}
\begin{figure}[h]
  \centering
  \includegraphics[width=0.95\linewidth]{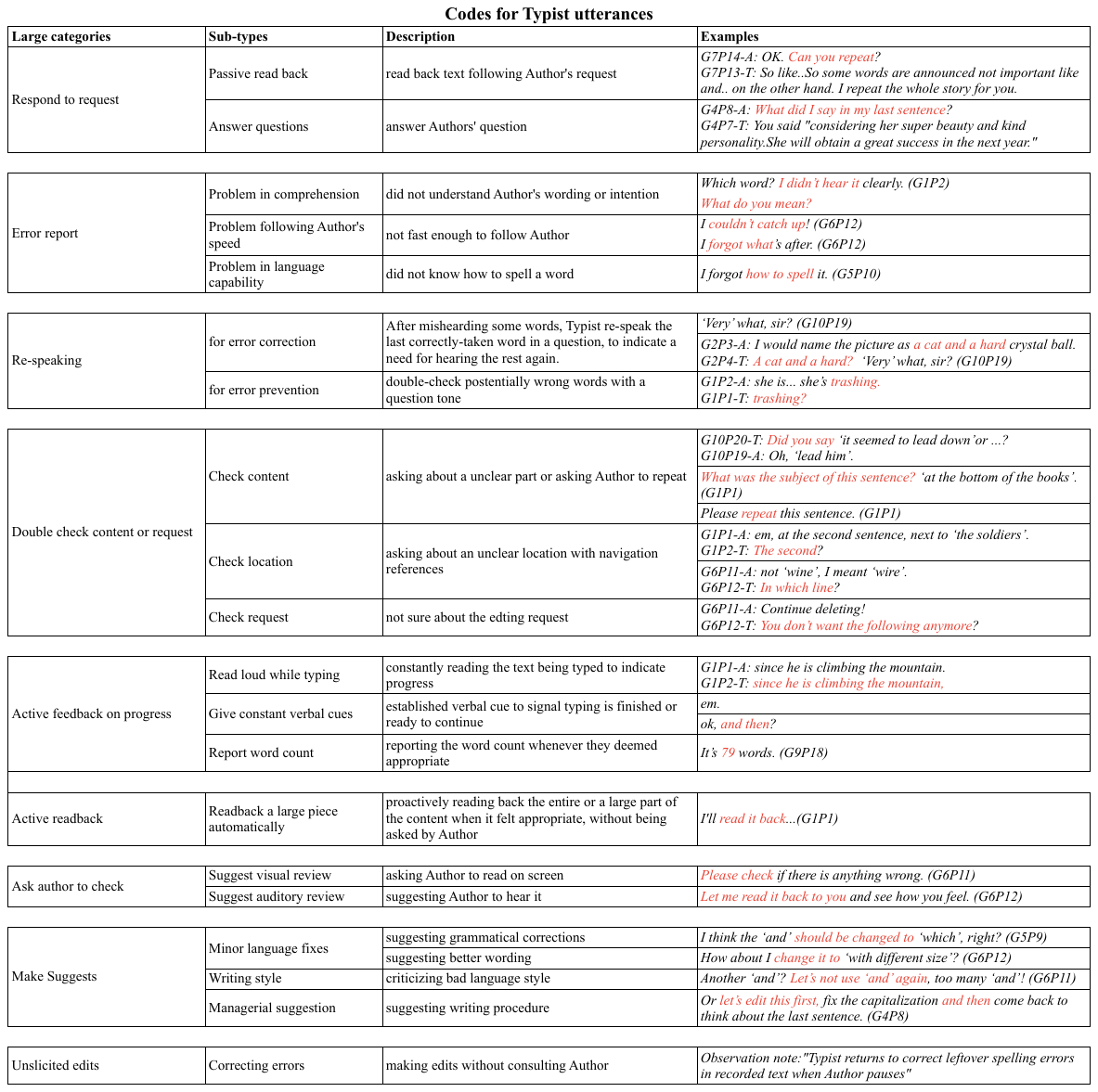}
\end{figure}

\end{document}